\documentclass[journal]{IEEEtran}

\usepackage{amsmath,amssymb,amsfonts}
\usepackage{amstext}
\usepackage{graphicx}
\usepackage{tikz}
\usetikzlibrary{arrows.meta, positioning}
\usepackage{tabularx}
\usepackage{makecell}
\usepackage{cite}
\usepackage{comment}
\usepackage{pgfplots}
\usepackage{algorithm}
\usepackage{algpseudocode}
\usepackage{multirow}

\pgfplotsset{compat=1.18}

\newtheorem{theorem}{Theorem}[section]
\newtheorem{definition}[theorem]{Definition}

\newtheorem{corollary}[theorem]{Corollary}
\newtheorem{remark}[theorem]{Remark}
\newtheorem{proposition}[theorem]{Proposition}
\newtheorem{assumption}{Assumption}

\begin{document}

\title{Copula-Based Dependence Modeling for Resilient 6G Multi-Hop Wireless Networks}

\author{
Ali~Izadimoein,~\IEEEmembership{Graduate Student Member,~IEEE,}
Prathapasinghe~Dharmawansa,
Nurul~Huda~Mahmood,~\IEEEmembership{Member,~IEEE,}
and~Eduard~Axel~Jorswieck,~\IEEEmembership{Fellow,~IEEE}
\thanks{Ali Izadimoein, Prathapasinghe Dharmawansa, and Nurul Huda Mahmood are with the Centre for Wireless Communications, University of Oulu, 90570 Oulu, Finland 
(e-mail: \{ali.izadimoein, prathapasinghe.kaluwadevage, nurulhuda.mahmood\}@oulu.fi).}
\thanks{Eduard Axel Jorswieck is with the Institute for Communications Technology, TU Braunschweig, 38106 Braunschweig, Germany 
(e-mail: e.jorswieck@tu-braunschweig.de).}
}


\maketitle

\begin{abstract}
Reliable operation of multi-hop wireless networks requires accounting for statistical dependence among link failures caused by common propagation conditions and shared network components. This paper develops a copula-based framework for modeling node-induced dependence among link-level signal-to-noise ratios and quantifying its effect on end-to-end outage. By separating marginal fading distributions from the dependence structure, the framework captures nonlinear and tail-dependent coupling beyond conventional correlation-based models. Exact outage expressions are derived for primary--backup configurations involving edge-disjoint paths and paths with shared links. We then characterize the conditional outage probability of each candidate backup path given a primary-path failure, jointly accounting for node-induced dependence and shared-link failures. In high-reliability regimes, we derive a first-order score that decomposes conditional outage into terms associated with marginal path outage, shared links, and node-induced dependence, enabling efficient dependence-aware backup-path selection. We also establish sufficient conditions under which the first-order and exact selection rules identify the same backup path. Numerical results validate the analysis and show that copula families with the same Kendall's \(\tau\) can yield markedly different outage behavior because of their different tail-dependence properties. In particular, lower-tail dependence and path overlap can substantially reduce redundancy gains and reverse rankings based only on marginal outage probability.
\end{abstract}

\IEEEpeerreviewmaketitle

\section{Introduction}
\label{sec:introduction}

Sixth-generation (6G) wireless communication technology is expected to transform connectivity by enabling ultra-high data rates, ultra-low latency, and massive device connectivity. These capabilities are particularly important for mission-critical applications, including industrial communication systems~\cite{popovski2018wireless}. Reliability and robustness are central design objectives for mission-critical applications that require uninterrupted operation. Reliability refers to a network's ability to operate consistently as intended and without failure, whereas robustness describes its ability to maintain acceptable performance under anticipated adverse conditions~\cite{kuo2003optimal}.

However, traditional reliability and robustness metrics are often insufficient in the presence of long-lasting or combined disruptions, such as consecutive packet losses, persistent interference, prolonged network outages, or adversarial attacks~\cite{sharkey2021search}. These metrics typically focus on short-term or isolated failures and may not adequately capture persistent or unexpected disruptions that can seriously reduce network performance~\cite{barker2013resilience}.

In contrast, resilience is a broader concept. It is not only about handling challenging conditions but also about anticipating problems, adapting to changes, and recovering from both expected and unexpected disruptions~\cite{mahmood2025resilient, alves20256gresiliencewhite}. This includes responding to extreme events such as natural disasters, cyberattacks, and sudden traffic surges~\cite{khaloopour2024resilience}. Specifically, resilience refers to a network's ability to maintain service continuity by either returning to a stable operational state or reconfiguring itself to operate under new conditions. The importance of resilience is especially evident in industrial networks and emergency scenarios and is expected to become even more critical in future 6G networks~\cite{berger2021survey}.

\subsection{Related Work}
\label{subsec:related_work}

Designing resilient wireless networks requires accurate reliability modeling and appropriate metrics for quantifying system performance. To assess communication reliability, several performance metrics have been proposed, among which outage probability is one of the most widely used~\cite{chaturvedi2016network}. At the physical layer, outage probability quantifies the likelihood that a communication link fails to satisfy a required quality threshold due to fading, interference, or limited radio resources~\cite{laneman2004cooperative}. 

Improving network reliability requires performance analysis that can accurately capture system behavior under varying operating conditions. In correlated fading scenarios, the joint cumulative distribution function approach has been widely used to account for spatial correlation or shared propagation paths~\cite{agrawal2009correlated,chen2004distribution}. However, this classical framework often assumes specific marginal distributions and characterizes channel dependence using linear correlation coefficients or covariance matrices~\cite{siwamogsatham2002new,shiu2000fading}. For instance, models for the joint distributions of correlated Rayleigh and Nakagami-$m$ fading gains have been extensively studied and are typically based on specific parametric formulations~\cite{mallik2003multivariate,karagiannidis2003multivariate}. Such assumptions are often violated in practical wireless environments, particularly in relay-assisted and multi-hop systems, where fading processes may exhibit nonlinear and asymmetric dependence structures~\cite{besser2021calculation,jorswieck2024massive}. These dependence structures affect joint outage behavior and reduce diversity gain. Accurate modeling of such effects is essential for reliable performance assessment under realistic conditions~\cite{gong2012interference}.

Despite its importance, accurately modeling statistical dependence in wireless networks remains a challenging problem. Most existing approaches rely on correlation-based metrics, which capture only linear and symmetric relationships. Such metrics characterize only a limited subset of the broader class of dependence structures and fail to reflect complex dependence behaviors observed in wireless channels~\cite{jorswieck2020copula,boche2004outage}. In particular, the same correlation value can correspond to different joint extreme-event probabilities under different dependence structures. Moreover, empirical studies show that wireless channels often exhibit asymmetric and nonlinear dependence~\cite{peters2014communications}.

To address these limitations, this paper adopts a copula-based framework that allows dependence structures to be modeled independently of marginal distributions. Copula theory has been widely used in finance, insurance, and risk management, and it has been increasingly applied to the performance analysis of wireless communication systems~\cite{besser2020copula,trigui2022copula,ghadi2022capacity,ghadi2020copula,besser2021fading}. These modeling approaches enable the characterization of the joint behavior of network components under realistic operating conditions, particularly in the analysis of joint outage events and dependent failures. Motivated by the success of copula models in other domains, this work aims to improve the analysis of dependent failures and facilitate a more accurate characterization of system resilience.

\subsection{Contributions}
\label{subsec:contributions}

Our preliminary work~\cite{izadimoein2025copula} used copulas to characterize statistical dependence between wireless fading links and to quantify its impact on link-level outage and redundancy. The present work advances that analysis to topology-aware multi-hop wireless networks. In particular, it models dependence between distinct cross-path links that share a common transmitter or receiver, distinguishes such statistical dependence from common failures caused by physically shared links, and develops a conditional-outage criterion for dependence-aware backup-path selection. The main contributions are summarized as follows:

\begin{itemize}

    \item We develop a copula-based dependence framework for multi-hop wireless networks in which statistical dependence may arise between cross-path links sharing a common transmitter or a common receiver. By separating the marginal signal-to-noise ratio distributions from the dependence structure, the framework captures nonlinear and tail-dependent coupling beyond conventional correlation-based models.

    \item We derive exact pairwise outage expressions for primary--backup multi-hop transmission under edge-disjoint and shared-link path configurations. The analysis distinguishes node-induced statistical dependence in non-shared cross-path link pairs from
    topology-induced common failures caused by links shared by the primary and backup paths.

    \item Under a pairwise block-factorized model in which the shared links, disjoint bivariate dependence blocks, and all remaining individual links are mutually independent, we derive the exact conditional outage probability of each candidate backup path given a failure of the primary path. The resulting expression separates the candidate path's marginal outage from the corrections caused by shared-link overlap and node-induced dependence.

    \item We develop a high-reliability first-order score that provides an additive decomposition of the conditional outage probability into contributions from marginal path outage, shared-link overlap, and node-induced dependence. Based on this score, we formulate a dependence-aware backup-path selection rule and establish sufficient conditions under which the first-order ranking agrees with the exact conditional-outage ranking.

    \item We provide numerical results that validate the analytical expressions and quantify the effects of copula family, dependence strength, lower-tail behavior, and shared-link topology. The results show that dependence structures with the same Kendall's \(\tau\) can produce substantially different outage behavior and that node-induced dependence and path overlap can reduce redundancy gains
    and reverse rankings based solely on marginal outage.

\end{itemize}

\noindent\textit{Notation:} The probability density function (PDF) and cumulative distribution function (CDF) of a random variable $X$ are denoted by $f_X(x)$ and $F_X(x)$, respectively. $\Pr(\cdot)$ denotes probability, and $\mathbb{E}[\cdot]$ denotes expectation. The Cartesian product of sets $\mathcal{A}$ and $\mathcal{B}$ is denoted by $\mathcal{A} \times \mathcal{B}$. The complement of an event $\mathcal{A}$ is denoted by $\mathcal{A}^c$. Vectors and matrices are denoted by bold lowercase and bold uppercase letters, respectively, e.g., $\mathbf{x}$ and $\mathbf{X}$. The symbols \(O(\cdot)\), \(o(\cdot)\), and \(\Theta(\cdot)\) denote standard asymptotic notation with respect to the stated limit. Specifically, \(f(x)=O(g(x))\) denotes an asymptotic upper bound, \(f(x)=o(g(x))\) means \(f(x)/g(x)\to0\), and \(f(x)=\Theta(g(x))\) means that \(c_1|g(x)|\leq|f(x)|\leq c_2|g(x)|\) for some constants \(c_1,c_2>0\) sufficiently near the stated limit.

\noindent The rest of this paper is organized as follows. Section~\ref{sec:system_model} presents the system model and formulates the outage and backup-path selection problems. Section~\ref{sec:copula_dependence} introduces the copula-based dependence framework, the copula families considered, and Kendall's $\tau$-based parameterization. Section~\ref{sec:outage_probability} derives outage probability expressions for edge-disjoint and shared-link path configurations and develops the conditional outage metric for backup-path selection. Section~\ref{sec:simulation} provides numerical results to validate the analysis and examine the effects of node-induced dependence and shared-link common failures. Finally, Section~\ref{sec:conclusion} concludes the paper.

\section{System Model and Problem Statement}
\label{sec:system_model}

This section presents the system model for multi-hop wireless transmission, including the network topology, channel model, and statistical dependence assumptions. The end-to-end reliability problem is then formulated in terms of outage probability and conditional outage metrics, leading to a dependence-aware backup-path selection problem.

\subsection{System Model}
\label{subsec:system_model}

We consider a wireless multi-hop network modeled as a directed graph \(G=(V,E)\), where \(V\) denotes a finite set of nodes and \(E=\{e_1,e_2,\ldots,e_n\}\subseteq V\times V\) denotes the set of directed communication links. Let \(s,d\in V\), \(s\neq d\), denote the source and destination nodes, respectively. A directed edge from node \(v_i\) to node \(v_j\) is denoted by \(e_{ij}=(v_i\to v_j)\). A directed path from \(s\) to \(d\) is denoted by \(\mathcal{P}_k\), has edge set \(E(\mathcal{P}_k)\), and consists of the directed-edge sequence \((v_0\to v_1),(v_1\to v_2),\ldots, (v_{L_k-1}\to v_{L_k})\), where \(v_0=s\) and \(v_{L_k}=d\). The wireless links are modeled according to the physical-layer characteristics of the corresponding channels. The proposed system model applies to general wireless multi-hop network topologies, and the set of candidate paths is assumed to be determined based on long-term channel statistics.

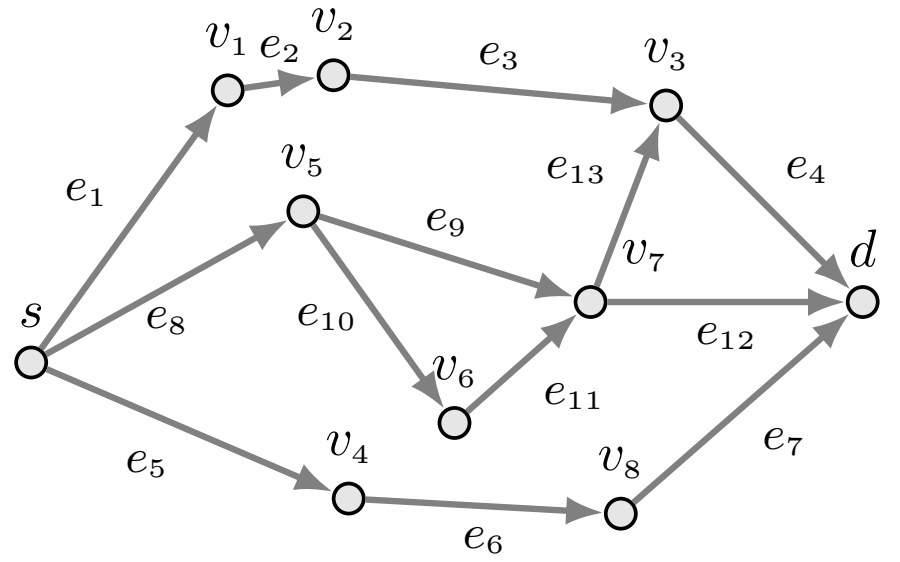
\begin{figure}[t]
\centering
\begin{tikzpicture}[
    scale=1,
    node/.style={circle, draw, line width=0.7pt, fill=gray!20, inner sep=2pt},
    base/.style={gray, line width=1.2pt, -{Latex[length=2.5mm]}}, 
    edgelabel/.style={font=\small, text=black}
]
\node[node, label=above:$s$]  (s)  at (0.7,-0.4) {};

\node[node, label=above:$v_{\scriptscriptstyle 1}$] (n1) at (2,1.4) {};
\node[node, label=above:$v_{\scriptscriptstyle 2}$] (n2) at (2.7,1.5) {};
\node[node, label=above:$v_{\scriptscriptstyle 3}$] (n3) at (4.9,1.3) {};

\node[node, label=above:$v_{\scriptscriptstyle 4}$] (n4) at (2.8,-1.3) {};
\node[node, label=above:$v_{\scriptscriptstyle 5}$] (n5) at (2.5,0.6) {};
\node[node, label=above:$v_{\scriptscriptstyle 6}$] (n6) at (3.5,-0.8) {};

\node[node, label=above right:$v_{\scriptscriptstyle 7}$] (n7) at (4.4,0) {};
\node[node, label=above:$v_{\scriptscriptstyle 8}$] (n8) at (4.6,-1.4) {};

\node[node, label=above:$d$]  (d)  at (6.2,0) {};

\draw[base] (s) -- node[above left, edgelabel] {$e_{\scriptscriptstyle 1}$} (n1);
\draw[base] (s) -- node[below left, edgelabel] {$e_{\scriptscriptstyle 5}$} (n4);

\draw[base] (n1) -- node[above, edgelabel] {$e_{\scriptscriptstyle 2}$} (n2);
\draw[base] (n2) -- node[above, edgelabel] {$e_{\scriptscriptstyle 3}$} (n3);

\draw[base] (s) -- node[below, edgelabel] {$e_{\scriptscriptstyle 8}$} (n5);
\draw[base] (n5) -- node[left, edgelabel] {$e_{\scriptscriptstyle 10}$} (n6);

\draw[base] (n5) -- node[above, edgelabel] {$e_{\scriptscriptstyle 9}$} (n7);
\draw[base] (n6) -- node[below right, edgelabel] {$e_{\scriptscriptstyle 11}$} (n7);

\draw[base] (n3) -- node[above right, edgelabel] {$e_{\scriptscriptstyle 4}$} (d);
\draw[base] (n7) -- node[below, edgelabel] {$e_{\scriptscriptstyle 12}$} (d);
\draw[base] (n8) -- node[below right, edgelabel] {$e_{\scriptscriptstyle 7}$} (d);

\draw[base] (n4) -- node[below, edgelabel] {$e_{\scriptscriptstyle 6}$} (n8);
\draw[base] (n7) -- node[above left, edgelabel] {$e_{\scriptscriptstyle 13}$} (n3);
\end{tikzpicture}
\caption{Example wireless network topology consisting of $10$ nodes and
$13$ directed communication links.}
\label{fig:network}
\vspace{-2mm}
\end{figure}

Fig.~\ref{fig:network} shows an example of the considered network, consisting of \(10\) nodes and \(13\) directed communication links. For instance, one path in Fig.~\ref{fig:network} has the edge set \(E(\mathcal{P}_1)=\{e_1,e_2,e_3,e_4\}\). Multiple paths may exist in the network. For example, \(E(\mathcal{P}_2)=\{e_5,e_6,e_7\}\) is edge-disjoint from \(E(\mathcal{P}_1)\), whereas \(E(\mathcal{P}_3)=\{e_8,e_9,e_{13},e_4\}\) shares the link \(e_4\) with \(E(\mathcal{P}_1)\), introducing a topology-induced common-failure effect between the corresponding path outage events. We define the sets of incoming and outgoing edges for any node \(v\in V\) as
\begin{equation}
\begin{aligned}
E_v^{\mathrm{in}} &= \{e_{uv}=(u,v)\in E \mid u\in V\}, \\
E_v^{\mathrm{out}} &= \{e_{vw}=(v,w)\in E \mid w\in V\}.
\end{aligned}
\end{equation}

Among the candidate paths, we designate \(\mathcal{P}_1\) as the primary path and \(\mathcal{P}_k\), \(k\geq 2\), as a candidate backup path. The primary path is assumed to be determined a priori using a conventional long-term routing criterion, such as minimum hop count, maximum average end-to-end SNR, or minimum marginal path outage probability. The focus of this work is therefore on evaluating and selecting backup paths given an outage of the primary path. The set of shared links between them is defined as
\begin{equation}
E_{1k}^{\mathrm{sh}}
\triangleq
E(\mathcal{P}_1)\cap E(\mathcal{P}_k).
\end{equation}

Moreover, let \(V_{1k}^{\mathrm{dep}}\) denote the set of nodes that induce statistical dependence between non-shared links of \(\mathcal{P}_1\) and \(\mathcal{P}_k\). For each such node, the relevant incoming or outgoing cross-path link pair is treated as one bivariate dependence block. The collection of these blocks is denoted by \(\mathcal{D}_{1k}\), and \(|\mathcal{D}_{1k}|\) represents the number of dependent link pairs considered between the two paths. A common internal node may contribute both an incoming and an outgoing dependence block; hence, \(|\mathcal{D}_{1k}|\) is not necessarily equal to \(|V_{1k}^{\mathrm{dep}}|\). Together, \(E_{1k}^{\mathrm{sh}}\) and \(\mathcal{D}_{1k}\) distinguish shared-link common failures from node-induced statistical dependence.

Link reliability is characterized using a signal-to-noise ratio (SNR) threshold. A link is considered successful if its instantaneous SNR exceeds the prescribed threshold corresponding to the target link quality. Accordingly, the instantaneous SNR at node \(v_j\) for the link from node \(v_i\) to node \(v_j\) is given by
\begin{equation}
\gamma_{ij} = \frac{P_t |h_{ij}|^2}{N_0 B d_{ij}^{\alpha}},
\label{eq:snr_instant}
\end{equation}
where \(P_t\) denotes the transmit power, \(N_0\) is the noise power spectral density, \(B\) is the system bandwidth, \(d_{ij}\) is the distance between nodes \(v_i\) and \(v_j\), and \(\alpha>2\) is the path-loss exponent. The complex channel coefficient \(h_{ij}\) captures small-scale fading. Although the framework can support different fading models, this paper focuses on Rayleigh fading for analytical tractability and numerical evaluation. The channel power gain is \(|h_{ij}|^2\), with \(\mathbb{E}[|h_{ij}|^2]=1\), corresponding to unit average channel power. Accordingly, the average SNR is given by
\begin{equation}
\bar{\gamma}_{ij} = \frac{P_t}{N_0 B d_{ij}^{\alpha}},
\label{eq:snr_avg}
\end{equation}
and the instantaneous SNR can be expressed as
\begin{equation}
\gamma_{ij} = \bar{\gamma}_{ij} |h_{ij}|^2.
\label{eq:snr_simple}
\end{equation}
For the directed communication link \(e_{ij}=(v_i \rightarrow v_j)\), a link outage occurs when the corresponding instantaneous SNR \(\gamma_{ij}\) falls below a prescribed threshold \(\gamma^{\mathrm{th}}_{ij}\). The resulting link outage probability is therefore defined as
\begin{equation*}
P_{ij}^{\mathrm{out}}
= \Pr\{\gamma_{ij} < \gamma^{\mathrm{th}}_{ij}\}.
\end{equation*}

We assume that links along an individual path are statistically independent, while dependence may arise within the incoming or outgoing link group of a common node. Thus, consecutive hops through the same relay, as well as pairs of links that do not share a common node, are treated as independent. This model is intended to represent dependence associated with terminal-side spatial correlation or common local scattering~\cite{shiu2000fading}, but excludes wider-area weather-induced dependence~\cite{yaghoubi2021design}. The relay-hop independence assumption is consistent with classical cooperative-relay models~\cite{laneman2004cooperative}; relay failure, congestion, common shadowing, or dependent consecutive hops would require an additional common-state model or a copula-based dual-hop model~\cite{ghadi2024performance}.

\subsection{Problem Statement}
\label{subsec:problem_statement}

We consider reliable data transmission from a source node \(s\) to a destination node \(d\) over a set of candidate multi-hop paths \(\mathcal{P}=\{\mathcal{P}_1,\mathcal{P}_2,\ldots,\mathcal{P}_K\}\), where \(L_k\) denotes the number of hops along path \(\mathcal{P}_k\). Relays are assumed to forward the message without error whenever the corresponding incoming and outgoing links are not in outage. Path selection is based on long-term channel statistics, such as average SNR, and instantaneous channel state information (CSI) is not available for adaptive path selection. Consequently, the considered path set remains fixed over each transmission interval, and an end-to-end outage occurs when all considered paths fail due to fading. If instantaneous CSI were available, adaptive path selection would lead to a different outage formulation. A path $\mathcal{P}_k$ is successful if all its links satisfy their
required SNR thresholds, i.e.,
\begin{equation*}
\mathcal{S}_k
=
\bigcap_{i=0}^{L_k-1}
\left\{
\gamma_{e_i^{(k)}}\geq
\gamma_{e_i^{(k)}}^{\mathrm{th}}
\right\},
\end{equation*}
where
$e_i^{(k)}=(v_i^{(k)}\to v_{i+1}^{(k)})$ denotes the $i$-th directed edge of $\mathcal{P}_k$, and $\gamma_{e_i^{(k)}}$ is its instantaneous SNR. Under the adopted independence assumption for links along an individual path, the path outage probability is
\begin{equation}
P_k^{\mathrm{out}}
=
\Pr(\mathcal{S}_k^c)
=
1-
\prod_{i=0}^{L_k-1}
\left(
1-P_{e_i^{(k)}}^{\mathrm{out}}
\right),
\label{eq:path_outage}
\end{equation}
where
\[
P_{e_i^{(k)}}^{\mathrm{out}}
=
\Pr\!\left(
\gamma_{e_i^{(k)}}<
\gamma_{e_i^{(k)}}^{\mathrm{th}}
\right).
\]

Among the available paths, $\mathcal{P}_1$ is designated as the primary path. It is assumed to be selected in advance using a conventional long-term routing criterion, such as minimum hop count, maximum average end-to-end SNR, or minimum marginal path outage probability. Since instantaneous CSI is not used for adaptive routing, the primary path remains fixed over the considered transmission interval. Let
\(\mathcal{B}=\{\mathcal{P}_2,\ldots,\mathcal{P}_K\}\)
denote the set of candidate backup paths from $s$ to $d$. The objective is to select a backup path that remains reliable when the primary path is in outage.

The suitability of a candidate backup path depends not only on its marginal outage probability but also on its tendency to fail together with the primary path. Accordingly, for each $\mathcal{P}_k\in\mathcal{B}$, we define the primary--candidate joint outage probability as
\begin{equation}
P_{1k}^{\mathrm{out}}
\triangleq
\Pr\!\left(
\mathcal{S}_1^c\cap\mathcal{S}_k^c
\right),
\qquad
k\in\{2,\ldots,K\}.
\label{eq:pairwise_joint_outage}
\end{equation}

Evaluating \eqref{eq:pairwise_joint_outage} requires characterizing the joint behavior of link SNRs associated with edges incident to common nodes, whereas shared links introduce topology-induced common-failure events. For notational convenience, let $\mathcal{E}_v\in\{E_v^{\mathrm{in}},E_v^{\mathrm{out}}\}$ denote one of the two link groups, and let $\mathcal{E}_v=\{e_1,\ldots,e_{|\mathcal{E}_v|}\}$ be an arbitrary ordering of its elements.
Let $\boldsymbol{\gamma}_{\mathcal{E}_v}$ and
$\boldsymbol{\gamma}_{\mathcal{E}_v}^{\mathrm{th}}$
denote the corresponding SNR and threshold vectors, respectively.
The joint CDF is
\begin{equation}
F_{\boldsymbol{\gamma}_{\mathcal{E}_v}}
\!\left(
\boldsymbol{\gamma}_{\mathcal{E}_v}^{\mathrm{th}}
\right)
=
\int_{0}^{\gamma_{e_1}^{\mathrm{th}}}
\cdots
\int_{0}^{\gamma_{e_{|\mathcal{E}_v|}}^{\mathrm{th}}}
f_{\boldsymbol{\gamma}_{\mathcal{E}_v}}
\!\left(
\boldsymbol{\gamma}_{\mathcal{E}_v}
\right)
\prod_{e\in\mathcal{E}_v}\mathrm{d}\gamma_e.
\label{eq:joint_snr_cdf}
\end{equation}
Direct specification of this joint distribution is generally difficult and may not yield tractable outage expressions. Moreover, a scalar correlation coefficient does not fully characterize nonlinear, asymmetric, or tail-dependent behavior governing joint low-SNR events. We therefore employ a copula-based framework that separates the marginal fading distributions from the dependence structure, thereby enabling flexible modeling of node-induced link dependence.

To quantify the reliability of candidate path \(\mathcal{P}_k\) after a failure of the primary path, we consider its conditional outage probability.
For $P_1^{\mathrm{out}}>0$,
\begin{equation}
\begin{aligned}
\Pr\!\left(
\mathcal{S}_k^c\mid\mathcal{S}_1^c
\right)
&=
\frac{
\Pr\!\left(
\mathcal{S}_1^c\cap\mathcal{S}_k^c
\right)
}{
\Pr\!\left(
\mathcal{S}_1^c
\right)
}
\\
&=
\frac{P_{1k}^{\mathrm{out}}}{P_1^{\mathrm{out}}}
=
P_k^{\mathrm{out}}
+
\Delta_{k\mid1},
\end{aligned}
\label{eq:cond_outage_general}
\end{equation}
where
\begin{equation}
\Delta_{k\mid1}
\triangleq
\Pr\!\left(
\mathcal{S}_k^c\mid\mathcal{S}_1^c
\right)
-
P_k^{\mathrm{out}}.
\label{eq:delta_cond_def_general}
\end{equation}
The term $\Delta_{k\mid1}$ measures the change in the outage probability of candidate path $\mathcal{P}_k$ caused by conditioning on a failure of the primary path. In particular, $\Delta_{k\mid1}=0$ when the two path-outage events are independent, whereas $\Delta_{k\mid1}>0$ indicates that their dependence reduces the available redundancy gain. This correction captures both node-induced statistical dependence and shared-link common-failure effects.

The conditional metric assumes a quasi-static network state. The primary path is attempted first, and, upon detection of its failure, the selected backup path is activated within the same fading realization. All link SNRs are assumed constant during primary-failure detection and backup transmission. Decode-and-forward relaying is employed, and the paths use orthogonal time or frequency resources so that inter-path interference is neglected. Switching errors, switching delay, and retransmission delay are not included.

The dependence-aware backup-path selection problem is therefore formulated as
\begin{equation}
\begin{aligned}
\mathcal{P}^{\star}
&\in
\arg\min_{\mathcal{P}_k\in\mathcal{B}}
\Pr\!\left(
\mathcal{S}_k^c\mid\mathcal{S}_1^c
\right),
\\
\mathcal{P}^{\star}
&\in
\arg\min_{\mathcal{P}_k\in\mathcal{B}}
\left(
P_k^{\mathrm{out}}
+
\Delta_{k\mid1}
\right).
\end{aligned}
\label{eq:backup_selection_conditional}
\end{equation}
Thus, the selected backup path accounts for both its marginal reliability and the dependence between its failure event and that of the primary path.

\section{Copula-Based Dependence Framework}
\label{sec:copula_dependence}

This section briefly reviews the copula tools used for dependence modeling, following~\cite{nelsen2006introduction}. Copulas separate marginal distributions from the dependence structure and are therefore suitable for modeling nonlinear and tail-dependent coupling beyond conventional correlation measures.

\begin{definition}[Copula]
A copula is a function \( C : [0,1]^q \to [0,1] \) that represents the joint cumulative distribution function of a \( q \)-dimensional random vector with standard uniform marginals.
\end{definition}

\begin{theorem}[Sklar's Theorem]
Let \(F_{\mathbf{X}}\) be the joint CDF of \(\mathbf{X}=(X_1,\ldots,X_q)\) with marginals \(F_{X_1},\ldots,F_{X_q}\). Then, there exists a copula \(C\) such that
\begin{equation}
F_{\mathbf{X}}(x_1,\ldots,x_q)
=
C\!\left(F_{X_1}(x_1),\ldots,F_{X_q}(x_q)\right).
\label{eq:copula_definition}
\end{equation}
If the marginals are continuous, \(C\) is unique.
\end{theorem}

For any copula \(C\) and \(\mathbf{u}=(u_1,\ldots,u_q)\in[0,1]^q\), the Fr\'echet--Hoeffding bounds are
\begin{equation}
W(\mathbf{u}) \le C(\mathbf{u}) \le M(\mathbf{u}),
\end{equation}
where
\begin{equation}
\begin{aligned}
W(\mathbf{u}) &= \max\left\{1-q+\sum_{i=1}^{q}u_i,\ 0\right\},\\
M(\mathbf{u}) &= \min(u_1,\ldots,u_q).
\end{aligned}
\end{equation}
Here, \(M\) is the Fr\'echet--Hoeffding upper-bound copula and represents comonotonic dependence. For \(q=2\), \(W\) is the Fr\'echet--Hoeffding lower-bound copula and represents countermonotonic dependence. For \(q>2\), \(W\) remains a pointwise lower bound for every \(q\)-dimensional copula, but it is not itself a valid copula.

\subsection{Copula Families and Dependence Measures}
\label{subsec:copula_families}

We consider representative Archimedean and elliptical copula families to capture symmetric and asymmetric dependence structures, including structures exhibiting tail dependence among wireless links.
\subsubsection{Clayton Copula}
\label{subsubsec:clayton_copula}
The Clayton copula captures lower-tail dependence. For \( \theta > 0 \), its \( q \)-dimensional form is
\begin{equation}
C_{\theta}^{\mathrm{Cl}}(\mathbf{u})
=
\left(
\sum_{i=1}^{q} u_i^{-\theta} - q + 1
\right)^{-\frac{1}{\theta}}.
\label{eq:clayton_copula}
\end{equation}

\subsubsection{Gumbel Copula}
\label{subsubsec:gumbel_copula}
The Gumbel copula captures upper-tail dependence. For \(\theta \ge 1\),
\begin{equation}
C_{\theta}^{\mathrm{Gu}}(\mathbf{u})
=
\exp \left(
-
\left[
\sum_{i=1}^{q}
\left( -\ln u_i \right)^{\theta}
\right]^{1/\theta}
\right).
\label{eq:gumbel_copula}
\end{equation}

\subsubsection{Gaussian Copula}
\label{subsubsec:gaussian_copula}
Let \(\mathbf{R}\) be a valid correlation matrix. The Gaussian copula is
\begin{equation}
C_{\mathbf{R}}^{\mathrm{Ga}}(\mathbf{u})
=
\Phi_{\mathbf{R}}\!\left(
\Phi^{-1}(u_1), \ldots, \Phi^{-1}(u_q)
\right),
\label{eq:gaussian_copula}
\end{equation}
where \(\Phi^{-1}(\cdot)\) is the inverse standard normal CDF and \(\Phi_{\mathbf{R}}(\cdot)\) is the multivariate normal CDF with correlation matrix \(\mathbf{R}\). The Gaussian copula captures symmetric dependence but has no tail dependence.

\subsubsection{Student-\(t\) Copula}
\label{subsubsec:student_t_copula}
The Student-\(t\) copula extends the Gaussian copula by capturing symmetric tail dependence. For a valid correlation matrix \(\mathbf{R}\) and a degrees-of-freedom parameter \(\nu>0\),
\begin{equation}
C_{\mathbf{R}, \nu}^{\mathrm{t}}(\mathbf{u}) 
=
t_{\mathbf{R}, \nu}\left(
t_\nu^{-1}(u_1), \ldots, t_\nu^{-1}(u_q)
\right),
\label{eq:student_copula}
\end{equation}
where \( t_\nu^{-1}(\cdot) \) is the inverse univariate Student-\(t\) CDF and \( t_{\mathbf{R}, \nu}(\cdot) \) is the corresponding multivariate CDF. As \( \nu \to \infty \), the Student-\(t\) copula approaches the Gaussian copula.

\subsubsection{Product Copula}
\label{subsubsec:product_copula}
The product copula is
\begin{equation}
C^{\perp}(\mathbf{u})
=
\prod_{i=1}^{q} u_i.
\label{eq:independence_copula}
\end{equation}
For continuous marginals, it characterizes independence.

\subsubsection{Dependence Measures}
To compare different copula families consistently, Kendall's \(\tau\) is used as a common dependence measure. For a bivariate random vector \((X,Y)\), it is defined as~\cite{nelsen2006introduction}
\begin{equation*}
\begin{aligned}
\tau &=
\Pr\!\big((X_1 - X_2)(Y_1 - Y_2) > 0\big) \\
&\quad -
\Pr\!\big((X_1 - X_2)(Y_1 - Y_2) < 0\big),
\end{aligned}
\end{equation*}
where $(X_1,Y_1)$ and $(X_2,Y_2)$ are independent realizations of $(X,Y)$. Fixing \(\tau\) allows different copula families to be compared at the same value of Kendall's \(\tau\). Table~\ref{tab:tau_relations} summarizes the parameter mappings used in the simulations.

\begin{table}[!t]
\centering
\caption{Kendall's \(\tau\)--based copula parameterization~\cite{nelsen2006introduction}.}
\label{tab:tau_relations}
\begin{tabularx}{\columnwidth}{|l|>{\centering\arraybackslash}X|}
\hline
\textbf{Copula Family} &
\textbf{Relationship Between $\tau$ and Parameters} \\
\hline
Gaussian/Student-\(t\) &
$\tau = \frac{2}{\pi}\arcsin(\rho)
\Rightarrow
\rho = \sin\!\left(\frac{\pi}{2}\tau\right)$ \\
\hline
Clayton &
$\tau = \frac{\theta}{\theta + 2}
\Rightarrow
\theta = \frac{2\tau}{1 - \tau}$ \\
\hline
Gumbel &
$\tau = 1 - \frac{1}{\theta}
\Rightarrow
\theta = \frac{1}{1 - \tau}$ \\
\hline
\end{tabularx}
\vspace{-2mm}
\end{table}

\section{Outage Probability Analysis and Backup-Path Selection}
\label{sec:outage_probability}

This section derives outage probability expressions for multi-hop wireless networks with node-induced statistical dependence. Copulas are employed to characterize the joint distributions of link SNRs, and the resulting formulations are developed for both edge-disjoint and shared-link path configurations. The analysis further yields a conditional outage metric for quantifying redundancy loss and enabling dependence-aware backup-path selection.

\subsection{Copula-Based Modeling of Joint Link SNR Distributions}
\label{subsec:copula_joint_snr}

As discussed in Section~\ref{subsec:problem_statement}, evaluating the system outage probability requires the joint distribution of link SNRs associated with links incident to common nodes. For a node \(v\), consider a link set \(\mathcal{E}_v \in \{E_v^{\mathrm{in}}, E_v^{\mathrm{out}}\}\). By Sklar's theorem~\eqref{eq:copula_definition}, the joint CDF in~\eqref{eq:joint_snr_cdf} can be written as
\begin{equation}
F_{\boldsymbol{\gamma}_{\mathcal{E}_v}}
\!\left(
\boldsymbol{\gamma}^{\mathrm{th}}_{\mathcal{E}_v}
\right)
=
C_{\mathcal{E}_v}\!\left(
F_{\gamma_{e_1}}(\gamma^{\mathrm{th}}_{e_1}),
\ldots,
F_{\gamma_{e_{|\mathcal{E}_v|}}}(\gamma^{\mathrm{th}}_{e_{|\mathcal{E}_v|}})
\right),
\label{eq:joint_cdf_copula}
\end{equation}
where \(C_{\mathcal{E}_v}(\cdot)\) is an \(|\mathcal{E}_v|\)-dimensional copula describing the dependence structure among the link SNRs in \(\mathcal{E}_v\). This representation separates the marginal SNR distributions from the dependence model, allowing different copula families to be assigned to different node-induced link groups.

For Rayleigh fading, the corresponding marginal outage probability is given by
$u_{e_i^{(k)}} =
F_{\gamma_{e_i^{(k)}}}
\!\left(
\gamma^{\mathrm{th}}_{e_i^{(k)}}
\right)
=
1-\exp(-\gamma^{\mathrm{th}}_{e_i^{(k)}}/\bar{\gamma}_{e_i^{(k)}})$. Other fading models can be incorporated by replacing the marginal CDFs accordingly.

\subsection{Outage Probability Analysis for Multiple-Path Networks}
\label{subsec:multi_path_outage}

In this subsection, we analyze the end-to-end outage probability in a multiple-path transmission setting. We first focus on two edge-disjoint paths between the source node \(s\) and the destination node \(d\), denoted by \(\mathcal{P}_1\) and \(\mathcal{P}_k\), where \(k\in\{2,\ldots,K\}\), such that \(E(\mathcal{P}_1)\cap E(\mathcal{P}_k)=\emptyset\). The analysis is then extended to candidate paths with shared links.

\begin{proposition}
\label{prop:exact_dual_path_outage}
Consider two edge-disjoint and internally node-disjoint paths \(\mathcal{P}_1\) and \(\mathcal{P}_k\), \(k\in\{2,\ldots,K\}\), from the source \(s\) to the destination \(d\), sharing only the endpoints \(s\) and \(d\). Let \(L_j\geq2\) for \(j\in\{1,k\}\). For path \(\mathcal{P}_j\), denote its first and last links by
\[
a_j\triangleq e_0^{(j)},
\qquad
b_j\triangleq e_{L_j-1}^{(j)}.
\]
For each link
\(e\in E(\mathcal{P}_1)\cup E(\mathcal{P}_k)\), define
\[
u_e
\triangleq
P_e^{\mathrm{out}}
=
F_{\gamma_e}\!\left(\gamma_e^{\mathrm{th}}\right).
\]

Assume that node-induced dependence exists only between the source-outgoing links \(a_1,a_k\) and between the destination-incoming links \(b_1,b_k\). Further, assume that the source-side and destination-side link groups and the intermediate links are mutually independent. Define
\[
\beta_j
\triangleq
\prod_{i=1}^{L_j-2}
\left(1-u_{e_i^{(j)}}\right),
\qquad
j\in\{1,k\},
\]
where an empty product is equal to one. Also define
\[
\Delta_s
\triangleq
C_{E_s^{\mathrm{out}}}
\left(u_{a_1},u_{a_k}\right)
-
u_{a_1}u_{a_k},
\]
and
\[
\Delta_d
\triangleq
C_{E_d^{\mathrm{in}}}
\left(u_{b_1},u_{b_k}\right)
-
u_{b_1}u_{b_k}.
\]

Then, the joint outage probability of the primary--candidate pair is
\begin{equation}
\begin{aligned}
P_{1k}^{\mathrm{out}}
&\triangleq
\Pr\!\left(
\mathcal{S}_1^c\cap\mathcal{S}_k^c
\right)
\\
&=
P_{1k,\mathrm{pind}}^{\mathrm{out}}
+
\beta_1\beta_k
\Big[
\Delta_s(1-u_{b_1})(1-u_{b_k})
\\
&\qquad\qquad
+
\Delta_d(1-u_{a_1})(1-u_{a_k})
+
\Delta_s\Delta_d
\Big],
\end{aligned}
\label{eq:exact_dual_path_outage_decomposition}
\end{equation}
where
\(P_{1k,\mathrm{pind}}^{\mathrm{out}}
= P_1^{\mathrm{out}}P_k^{\mathrm{out}}\)
is the path-independent benchmark, with
\[
P_j^{\mathrm{out}}
=
1-(1-u_{a_j})\beta_j(1-u_{b_j}),
\qquad
j\in\{1,k\}.
\]
\end{proposition}

The proof of Proposition~\ref{prop:exact_dual_path_outage} is provided in Appendix~\ref{app:exact_dual_path_outage}.

For \(v\in\{s,d\}\), let \((e_1,e_2)\) denote the relevant cross-path link pair in \(\mathcal{E}_v\). Under Rayleigh fading, the dependence correction
\(\Delta_v\) can be evaluated in closed form for the bivariate Clayton and Gumbel copulas by substituting the marginal Rayleigh outage probabilities into \eqref{eq:clayton_copula} and \eqref{eq:gumbel_copula}, respectively. The product copula yields \(\Delta_v=0\). For a Clayton copula with parameter
\(\theta_v\), the dependence correction is
\begin{equation}
\begin{aligned}
\Delta_v
&=
\left(
\sum_{i=1}^{2}
\left[
1 - \exp\!\left(-\frac{\gamma_{e_i}^{\mathrm{th}}}{\bar{\gamma}_{e_i}}\right)
\right]^{-\theta_v}
- 1
\right)^{-1/\theta_v}
\\
&\quad -
\prod_{i=1}^{2}
\left[
1 - \exp\!\left(-\frac{\gamma_{e_i}^{\mathrm{th}}}{\bar{\gamma}_{e_i}}\right)
\right].
\end{aligned}
\end{equation}
For a Gumbel copula with parameter $\theta_v$, the dependence correction is
\begin{equation}
\begin{aligned}
\Delta_v
&=
\exp\!\left(
-
\left[
\sum_{i=1}^{2}
\left(
-\ln
\left(
1 - \exp\!\left(
-\frac{\gamma_{e_i}^{\mathrm{th}}}
{\bar{\gamma}_{e_i}}
\right)
\right)
\right)^{\theta_v}
\right]^{1/\theta_v}
\right)
\\
&\quad -
\prod_{i=1}^{2}
\left[
1 - \exp\!\left(
-\frac{\gamma_{e_i}^{\mathrm{th}}}
{\bar{\gamma}_{e_i}}
\right)
\right].
\end{aligned}
\end{equation}
For the Gaussian and Student-\(t\) copulas, analogous expressions for \(\Delta_v\) can be obtained, but they generally do not admit closed-form representations and are evaluated numerically.

For edge-disjoint primary--candidate paths, substituting
\eqref{eq:exact_dual_path_outage_decomposition} into
\eqref{eq:delta_cond_def_general} gives
\begin{equation}
\begin{aligned}
\Delta_{k\mid1}
&=
\frac{\beta_1\beta_k}{P_1^{\mathrm{out}}}
\Big[
\Delta_s(1-u_{b_1})(1-u_{b_k})
\\
&\qquad
+
\Delta_d(1-u_{a_1})(1-u_{a_k})
+
\Delta_s\Delta_d
\Big].
\end{aligned}
\label{eq:conditional_outage_gain_exact}
\end{equation}
Here, \(\Delta_s\) and \(\Delta_d\) denote the dependence corrections associated with the source-side pair \((a_1,a_k)\) and the destination-side pair \((b_1,b_k)\), respectively. A positive value of \(\Delta_{k\mid1}\) indicates that node-induced dependence increases the conditional outage probability of candidate path \(\mathcal{P}_k\) relative to its marginal outage probability, thereby reducing the redundancy gain.

\begin{corollary}
\label{cor:high_reliability_dual_path}
Consider the edge-disjoint primary--candidate path configuration in Proposition~\ref{prop:exact_dual_path_outage}. Let
\[
u_{\max}
\triangleq
\max_{e\in E(\mathcal{P}_1)\cup E(\mathcal{P}_k)}u_e,
\]
and assume that \(u_{\max}\to0\), the path lengths remain finite, and the source-side and destination-side link pairs exhibit positive quadrant dependence. Then,
\begin{equation}
P_{1k}^{\mathrm{out}}
=
P_{1k,\mathrm{pind}}^{\mathrm{out}}
+
\Delta_s+\Delta_d
+
O\!\left(
u_{\max}(\Delta_s+\Delta_d)
\right).
\label{eq:pairwise_outage_high_reliability}
\end{equation}
Hence, to leading order,
\begin{equation}
P_{1k}^{\mathrm{out}}
\approx
P_{1k,\mathrm{pind}}^{\mathrm{out}}
+
\Delta_s+\Delta_d.
\label{eq:pairwise_outage_approx}
\end{equation}
Moreover,
\begin{equation}
\Delta_{k\mid1}
=
\frac{\Delta_s+\Delta_d}{P_1^{\mathrm{out}}}
\left[1+O(u_{\max})\right]
\approx
\frac{\Delta_s+\Delta_d}{P_1^{\mathrm{out}}},
\label{eq:delta_k_given_1_approx}
\end{equation}
provided that \(\Delta_s+\Delta_d>0\). If
\(\Delta_s=\Delta_d=0\), then \(\Delta_{k\mid1}=0\).
\end{corollary}

\noindent\textit{Exact pairwise outage with shared links:}
The following factorization is based on the node-local dependence model and the relay-hop independence assumption described in Section~\ref{subsec:system_model}.
\begin{assumption}[Block-factorized cross-path dependence]
\label{ass:block_factorized_dependence}
For each primary--candidate path pair \((\mathcal{P}_1,\mathcal{P}_k)\), the non-shared links exhibiting cross-path dependence can be partitioned into a collection \(\mathcal{D}_{1k}\) of disjoint bivariate dependence blocks. Each block \(b\in\mathcal{D}_{1k}\) consists of one non-shared link \(e_{b,1}\in E(\mathcal{P}_1)\setminus E_{1k}^{\mathrm{sh}}\) and one non-shared link \(e_{b,k}\in E(\mathcal{P}_k)\setminus E_{1k}^{\mathrm{sh}}\), and no link belongs to more than one block. The dependence blocks, the distinct shared links in \(E_{1k}^{\mathrm{sh}}\), and all remaining individual links are mutually independent. Within each bivariate block, dependence is characterized by the corresponding copula \(C_{b,k}\), while each shared link is treated as a single common random variable appearing in both paths.
\end{assumption}
\begin{remark}
A group \(\mathcal{E}_g\) containing three or more dependent links can be modeled by assigning an \(|\mathcal{E}_g|\)-dimensional copula \(C_{\mathcal{E}_g}\) to their marginal SNR distributions through Sklar's theorem. The corresponding joint link-success probability can then be obtained from the resulting multivariate CDF by inclusion--exclusion. For analytical tractability, the exact derivations in this paper focus on disjoint bivariate dependence blocks that are mutually independent.
\end{remark}
In what follows, ``exact'' refers to evaluation under Assumption~\ref{ass:block_factorized_dependence}. Under this assumption, define the marginal path-success probabilities as
\begin{equation}
\zeta_j
\triangleq
\Pr(\mathcal{S}_j)
=
\prod_{e\in E(\mathcal{P}_j)}(1-u_e)
=
1-P_j^{\mathrm{out}},
\qquad j\in\{1,k\}.
\label{eq:path_success_zeta}
\end{equation}
For each dependence block \(b\in\mathcal{D}_{1k}\), let \(e_{b,1}\in E(\mathcal{P}_1)\) and \(e_{b,k}\in E(\mathcal{P}_k)\) denote the corresponding non-shared link pair, and define
\begin{equation}
\Delta_{b,k}
\triangleq
C_{b,k}
\left(
u_{e_{b,1}},u_{e_{b,k}}
\right)
-
u_{e_{b,1}}u_{e_{b,k}}.
\label{eq:block_dependence_correction_exact}
\end{equation}
The shared-link and node-induced dependence factors are
\begin{equation}
\Lambda_{1k}^{\mathrm{sh}}
\triangleq
\prod_{e\in E_{1k}^{\mathrm{sh}}}
\frac{1}{1-u_e},
\label{eq:exact_shared_factor}
\end{equation}
and
\begin{equation}
\Lambda_{1k}^{\mathrm{dep}}
\triangleq
\prod_{b\in\mathcal{D}_{1k}}
\left(
1+
\frac{
\Delta_{b,k}
}{
(1-u_{e_{b,1}})(1-u_{e_{b,k}})
}
\right),
\label{eq:exact_dependence_factor}
\end{equation}
where an empty product is equal to one. The exact joint path-success probability is then
\begin{equation}
\Pr(\mathcal{S}_1\cap\mathcal{S}_k)
=
\zeta_1\zeta_k
\Lambda_{1k}^{\mathrm{sh}}
\Lambda_{1k}^{\mathrm{dep}}.
\label{eq:exact_joint_path_success}
\end{equation}
Consequently, the exact primary--candidate joint outage probability is
\begin{equation}
\begin{aligned}
P_{1k}^{\mathrm{out}}
&\triangleq
\Pr(\mathcal{S}_1^c\cap\mathcal{S}_k^c)
\\
&=
1-\zeta_1-\zeta_k
+
\zeta_1\zeta_k
\Lambda_{1k}^{\mathrm{sh}}
\Lambda_{1k}^{\mathrm{dep}}
\\
&=
P_{1k,\mathrm{pind}}^{\mathrm{out}}
+
\zeta_1\zeta_k
\left(
\Lambda_{1k}^{\mathrm{sh}}
\Lambda_{1k}^{\mathrm{dep}}
-1
\right),
\end{aligned}
\label{eq:exact_shared_path_outage}
\end{equation}
where
\(P_{1k,\mathrm{pind}}^{\mathrm{out}}
= P_1^{\mathrm{out}}P_k^{\mathrm{out}}\)
is the path-independent benchmark for the primary--candidate pair.

Removing node-induced dependence while preserving the shared-link topology gives the pairwise node-independent benchmark
\begin{equation}
P_{1k,\mathrm{ind}}^{\mathrm{out}}
=
1-\zeta_1-\zeta_k
+
\zeta_1\zeta_k
\Lambda_{1k}^{\mathrm{sh}}.
\label{eq:exact_shared_node_independent}
\end{equation}
Hence, the exact pairwise outage probability can equivalently be written as
\begin{equation}
P_{1k}^{\mathrm{out}}
=
P_{1k,\mathrm{ind}}^{\mathrm{out}}
+
\zeta_1\zeta_k
\Lambda_{1k}^{\mathrm{sh}}
\left(
\Lambda_{1k}^{\mathrm{dep}}-1
\right).
\label{eq:exact_shared_outage_decomposition}
\end{equation}
The second term in
\eqref{eq:exact_shared_outage_decomposition} is the exact correction caused by node-induced dependence, whereas the shared-link effect is retained in
\(P_{1k,\mathrm{ind}}^{\mathrm{out}}\).

Under Assumption~\ref{ass:block_factorized_dependence} and for \(P_1^{\mathrm{out}}>0\), the exact conditional outage probability of candidate path \(\mathcal{P}_k\) is
\begin{equation}
\begin{aligned}
\Pr(\mathcal{S}_k^c\mid\mathcal{S}_1^c)
&=
\frac{P_{1k}^{\mathrm{out}}}{P_1^{\mathrm{out}}}
=
P_k^{\mathrm{out}}
+
\frac{\zeta_1\zeta_k}{P_1^{\mathrm{out}}}
\left(
\Lambda_{1k}^{\mathrm{sh}}
\Lambda_{1k}^{\mathrm{dep}}
-1
\right).
\end{aligned}
\label{eq:exact_conditional_outage_blocks}
\end{equation}
Therefore, the exact conditional correction in
\eqref{eq:delta_cond_def_general} is
\begin{equation}
\begin{aligned}
\Delta_{k\mid1}
&=
\frac{\zeta_1\zeta_k}{P_1^{\mathrm{out}}}
\Big[
\left(\Lambda_{1k}^{\mathrm{sh}}-1\right)
+
\Lambda_{1k}^{\mathrm{sh}}
\left(\Lambda_{1k}^{\mathrm{dep}}-1\right)
\Big].
\end{aligned}
\label{eq:exact_conditional_correction_blocks}
\end{equation}
The first term in
\eqref{eq:exact_conditional_correction_blocks} is the exact topology-induced correction caused by shared links, while the second
term is the additional correction caused by node-induced dependence.

\subsection{Dependence-Aware Backup-Path Selection}
\label{subsec:path_selection}

The conditional outage probability in \eqref{eq:exact_conditional_outage_blocks} provides the criterion for dependence-aware backup-path selection.

\subsubsection*{Algorithmic Implementation}

Under Assumption~\ref{ass:block_factorized_dependence} and for $P_1^{\mathrm{out}}>0$, the exact selection rule is implemented using Algorithm~\ref{alg:backup_selection}.

\begin{algorithm}[t]
\caption{Exact dependence-aware backup-path selection}
\label{alg:backup_selection}
\begin{algorithmic}[1]
\Require Primary path $\mathcal{P}_1$, candidate set $\mathcal{B}$,
link outage probabilities $u_e$, shared-link sets
$E_{1k}^{\mathrm{sh}}$, dependence-block sets $\mathcal{D}_{1k}$,
and associated bivariate copulas
\Ensure Selected backup path $\mathcal{P}^{\star}$

\State Compute $P_1^{\mathrm{out}}$ using \eqref{eq:path_outage}
\State $Q_{\min}\gets\infty$

\For{each $\mathcal{P}_k\in\mathcal{B}$}
    \State Compute $P_k^{\mathrm{out}}$ using \eqref{eq:path_outage}
    \State Compute $\Delta_{k\mid1}$ using
    \eqref{eq:exact_conditional_correction_blocks}
    \State $Q_k\gets P_k^{\mathrm{out}}+\Delta_{k\mid1}$
    \If{$Q_k<Q_{\min}$}
        \State $Q_{\min}\gets Q_k$
        \State $\mathcal{P}^{\star}\gets\mathcal{P}_k$
    \EndIf
\EndFor

\State \Return $\mathcal{P}^{\star}$
\end{algorithmic}
\end{algorithm}
Candidates are processed in increasing order of path index. If multiple candidates attain the same value of the exact score \(Q_k\), the candidate with the smallest path index is selected.

Assuming that the shared-link and dependence-block sets are precomputed and that each bivariate copula evaluation requires $O(1)$ operations to achieve a prescribed numerical accuracy, the computational complexity is
\begin{equation}
O\!\left(
L_1+
\sum_{\mathcal{P}_k\in\mathcal{B}}
\left[
L_k+
|E_{1k}^{\mathrm{sh}}|+
|\mathcal{D}_{1k}|
\right]
\right).
\label{eq:algorithm_complexity}
\end{equation}
Computing the primary-path outage requires $O(L_1)$ operations. For each candidate $\mathcal{P}_k$, evaluating its marginal outage, shared-link contribution, and node-induced dependence contribution requires $O(L_k)$, $O(|E_{1k}^{\mathrm{sh}}|)$, and $O(|\mathcal{D}_{1k}|)$ operations, respectively. The $O(|\mathcal{B}|)$ comparisons are absorbed into \eqref{eq:algorithm_complexity}. This complexity excludes candidate-path generation and construction of the shared-link and dependence-block sets.

\subsection{High-Reliability Approximation for Backup-Path Selection}
\label{subsec:high_reliability_backup_selection}

Although the exact conditional outage probability can be used for backup-path selection, a high-reliability approximation provides an additive decomposition into contributions from marginal outage, shared links, and node-induced dependence.

For each candidate path \(\mathcal{P}_k\in\mathcal{B}\), define
\begin{equation}
\begin{aligned}
\Omega_k^{\mathrm{sh}}
&\triangleq
\sum_{e\in E_{1k}^{\mathrm{sh}}}u_e,
&
\Omega_k^{\mathrm{dep}}
&\triangleq
\sum_{b\in\mathcal{D}_{1k}}\Delta_{b,k},
\\
A_k
&\triangleq
\Omega_k^{\mathrm{sh}}
+
\Omega_k^{\mathrm{dep}}.
\end{aligned}
\label{eq:omega_definitions}
\end{equation}
Here, \(\Omega_k^{\mathrm{sh}}\) and \(\Omega_k^{\mathrm{dep}}\) are the first-order shared-link and node-induced dependence contributions, respectively, and \(\Delta_{b,k}\) is defined in
\eqref{eq:block_dependence_correction_exact}. Define
\begin{equation}
u_{\max}
\triangleq
\max_{
e\in
E(\mathcal{P}_1)
\cup
\left(
\displaystyle\bigcup_{\mathcal{P}_j\in\mathcal{B}}
E(\mathcal{P}_j)
\right)}
u_e.
\label{eq:backup_selection_umax}
\end{equation}
We consider the high-reliability regime \(u_{\max}\to0\), while the candidate set, path topologies, path lengths, copula parameters, and numbers of shared links and dependence blocks remain fixed. We further assume that at least one link of the primary path has outage probability \(\Theta(u_{\max})\). Therefore,
\begin{equation}
P_1^{\mathrm{out}}
=
\Theta(u_{\max}),
\qquad
P_k^{\mathrm{out}}
=
O(u_{\max}),
\quad
\mathcal{P}_k\in\mathcal{B}.
\label{eq:path_outage_scaling}
\end{equation}
For the high-reliability analysis, each dependence block is assumed to satisfy \(C_{b,k}(u_{e_{b,1}},u_{e_{b,k}}) \geq u_{e_{b,1}}u_{e_{b,k}}\), and hence \(\Delta_{b,k}\geq0\). This condition holds globally for the product copula (with equality), the Clayton copula with \(\theta>0\), the Gumbel copula with \(\theta\geq1\), and the Gaussian copula with \(\rho\geq0\) (equivalently, \(\tau\geq0\))~\cite{nelsen2006introduction}. For the finite-\(\nu\) Student-\(t\) copula, this condition is imposed at the relevant outage-probability pairs. Under this condition, the Fr\'echet--Hoeffding bound implies \(\Delta_{b,k}=O(u_{\max})\). Since the numbers of shared links and dependence blocks remain fixed,
\begin{equation}
\Omega_k^{\mathrm{sh}}
=
O(u_{\max}),
\qquad
\Omega_k^{\mathrm{dep}}
=
O(u_{\max}),
\qquad
A_k
=
O(u_{\max}).
\label{eq:aggregate_correction_scaling}
\end{equation}

\begin{proposition}
\label{prop:first_order_conditional_score}
Under Assumption~\ref{ass:block_factorized_dependence} and the nonnegative block-correction condition above, consider the high-reliability regime \(u_{\max}\to0\) with a fixed finite candidate set \(\mathcal{B}\). Then, for every \(\mathcal{P}_k\in\mathcal{B}\),
\begin{equation}
\Lambda_{1k}^{\mathrm{sh}}
\Lambda_{1k}^{\mathrm{dep}}
-1
=
A_k
+
O\!\left(
u_{\max}A_k
\right).
\label{eq:combined_factor_first_order}
\end{equation}
Consequently, for \(P_1^{\mathrm{out}}>0\),
\begin{equation}
\begin{aligned}
\Pr\!\left(
\mathcal{S}_k^c\mid\mathcal{S}_1^c
\right)
&=
P_k^{\mathrm{out}}
+
\frac{A_k}{P_1^{\mathrm{out}}}
+
O\!\left(
\frac{
u_{\max}A_k
}{
P_1^{\mathrm{out}}
}
\right)
\\
&=
P_k^{\mathrm{out}}
+
\frac{
\Omega_k^{\mathrm{sh}}
+
\Omega_k^{\mathrm{dep}}
}{
P_1^{\mathrm{out}}
}
+
O(A_k),
\end{aligned}
\label{eq:conditional_outage_score_expansion}
\end{equation}
where the second equality follows from
\(A_k=\Omega_k^{\mathrm{sh}}+\Omega_k^{\mathrm{dep}}\) and \(P_1^{\mathrm{out}}=\Theta(u_{\max})\).
\end{proposition}

The proof of Proposition~\ref{prop:first_order_conditional_score}
is provided in Appendix~\ref{app:first_order_conditional_score}.

\subsubsection*{First-Order Backup-Path Ranking}
\label{subsubsec:first_order_backup_ranking}

Based on Proposition~\ref{prop:first_order_conditional_score}, define the first-order score of candidate path \(\mathcal{P}_k\) as
\begin{equation}
J_k
\triangleq
P_k^{\mathrm{out}}
+
\frac{A_k}{P_1^{\mathrm{out}}}
=
P_k^{\mathrm{out}}
+
\frac{
\Omega_k^{\mathrm{sh}}+\Omega_k^{\mathrm{dep}}
}{
P_1^{\mathrm{out}}
}.
\label{eq:ranking_metric_jk}
\end{equation}
If multiple candidates attain the same first-order score \(J_k\), the candidate
with the smallest path index is selected.

Then,
\begin{equation}
\Pr(\mathcal{S}_k^c\mid\mathcal{S}_1^c)
=
J_k
+
O\!\left(
\frac{u_{\max}A_k}{P_1^{\mathrm{out}}}
\right)
=
J_k+O(A_k),
\label{eq:conditional_outage_score_jk}
\end{equation}
where \(P_1^{\mathrm{out}}=\Theta(u_{\max})\). Thus, \(J_k\) combines the exact marginal outage probability with first-order shared-link and node-induced dependence corrections. If \(A_k=0\), then \(J_k=P_k^{\mathrm{out}}\), which equals the exact conditional outage probability.

Accordingly, the proposed first-order backup-path selection rule is
\begin{equation}
\widehat{\mathcal{P}}^{\star}
\in
\arg\min_{\mathcal{P}_k\in\mathcal{B}} J_k.
\label{eq:backup_selection_optimization}
\end{equation}
The score \(J_k\) provides a high-reliability approximation to the exact conditional outage probability \(Q_k\). Its additive form separates the effects of marginal path outage, shared-link overlap, and node-induced statistical dependence, thereby facilitating the interpretation and comparison of candidate backup paths.

From Proposition~\ref{prop:first_order_conditional_score},
\begin{equation}
Q_k
=
J_k
+
O\!\left(
\frac{u_{\max}A_k}{P_1^{\mathrm{out}}}
\right).
\label{eq:first_order_score_accuracy}
\end{equation}
Hence, \(J_k\) approaches \(Q_k\) for each candidate; however, ranking agreement additionally requires sufficient separation between candidate scores. When candidate scores are close or the link outage probabilities are not sufficiently small, the exact metric \(Q_k\) should be used for backup-path selection.

\subsubsection*{Dependence-Aware Path Ranking}
\label{subsubsec:design_insights}

For any two candidate backup paths \(\mathcal{P}_i\) and
\(\mathcal{P}_j\), the difference between their first-order
scores is
\begin{equation}
\begin{aligned}
J_i-J_j
&=
P_i^{\mathrm{out}}-P_j^{\mathrm{out}}
\\
&\quad+
\frac{
\Omega_i^{\mathrm{sh}}+\Omega_i^{\mathrm{dep}}
-\Omega_j^{\mathrm{sh}}-\Omega_j^{\mathrm{dep}}
}{
P_1^{\mathrm{out}}
}.
\end{aligned}
\label{eq:general_score_difference}
\end{equation}
Suppose that \(P_i^{\mathrm{out}}>P_j^{\mathrm{out}}\). Then,
\(\mathcal{P}_i\) is ranked ahead of \(\mathcal{P}_j\) by the
first-order score, i.e., \(J_i<J_j\), if and only if
\begin{equation}
\begin{aligned}
&\Omega_j^{\mathrm{sh}}+\Omega_j^{\mathrm{dep}}
-\Omega_i^{\mathrm{sh}}-\Omega_i^{\mathrm{dep}}
>
P_1^{\mathrm{out}}
\left(
P_i^{\mathrm{out}}-P_j^{\mathrm{out}}
\right).
\end{aligned}
\label{eq:marginal_outage_reversal}
\end{equation}
A candidate with a higher marginal outage probability may still be preferred if it has smaller shared-link and dependence penalties. Thus, the path with the best stand-alone reliability is not always the best backup. Under the adopted model, a candidate whose failure event is less dependent on that of the primary path may be preferred despite having a higher marginal outage probability.

For an edge-disjoint candidate, \(\Omega_k^{\mathrm{sh}}=0\); otherwise,
\(\Omega_k^{\mathrm{sh}}=\sum_{e\in E_{1k}^{\mathrm{sh}}}u_e\).
When two candidates have the same marginal outage probability and node-induced dependence contribution, the edge-disjoint path is preferred, with a strict preference if the other candidate has \(\Omega_k^{\mathrm{sh}}>0\). Within the adopted model, shared links increase the conditional-outage score. This effect must be balanced against any marginal-outage or hop-count disadvantage associated with an edge-disjoint candidate.

For two candidates that are edge-disjoint from the primary path and satisfy \(P_i^{\mathrm{out}}>P_j^{\mathrm{out}}\), \eqref{eq:marginal_outage_reversal} reduces to
\begin{equation}
\Omega_j^{\mathrm{dep}}-\Omega_i^{\mathrm{dep}}
>
P_1^{\mathrm{out}}
\left(
P_i^{\mathrm{out}}-P_j^{\mathrm{out}}
\right).
\label{eq:disjoint_outage_reversal}
\end{equation}
Since
\[
P_1^{\mathrm{out}}=\Theta(u_{\max}),
\qquad
P_i^{\mathrm{out}}-P_j^{\mathrm{out}}=O(u_{\max}),
\]
the right-hand side of
\eqref{eq:disjoint_outage_reversal} is \(O(u_{\max}^2)\).
Therefore, the sufficient condition
\begin{equation}
\frac{
\Omega_j^{\mathrm{dep}}-\Omega_i^{\mathrm{dep}}
}{
u_{\max}^2
}
\longrightarrow +\infty,
\label{eq:dependence_dominance_condition}
\end{equation}
ensures that the dependence advantage of \(\mathcal{P}_i\) asymptotically dominates its marginal-outage disadvantage. This is especially relevant when cross-path links are incident to the same terminal or experience similar local propagation conditions. 

If all dependence blocks associated with candidate
\(\mathcal{P}_k\) have the same correction
\(\Delta_{\mathrm{eq},k}\), then
\begin{equation}
\Omega_k^{\mathrm{dep}}
=
|\mathcal{D}_{1k}|\Delta_{\mathrm{eq},k}.
\label{eq:equal_dependence_corrections}
\end{equation}
Otherwise, the block-specific corrections must be summed. Thus, the aggregate dependence contribution reflects both the number of dependent cross-path link pairs and the strength of dependence within each pair.

Strong lower-tail dependence can generate a nonvanishing conditional-outage penalty in the high-reliability regime.
Suppose that \(\Omega_i^{\mathrm{dep}}=0\) and that, for some block \(b\in\mathcal{D}_{1j}\),
\begin{equation}
u_{e_{b,1}}=\Theta(u_{\max}),
\qquad
u_{e_{b,j}}=\Theta(u_{\max}),
\label{eq:tail_block_marginal_scaling}
\end{equation}
and
\begin{equation}
C_{b,j}\!\left(
u_{e_{b,1}},u_{e_{b,j}}
\right)
=
\Theta(u_{\max}).
\label{eq:tail_block_copula_scaling}
\end{equation}
Because
\[
u_{e_{b,1}}u_{e_{b,j}}
=
\Theta(u_{\max}^2),
\]
the corresponding dependence correction satisfies
\begin{equation}
\begin{aligned}
\Delta_{b,j}
&=
C_{b,j}\!\left(
u_{e_{b,1}},u_{e_{b,j}}
\right)
-u_{e_{b,1}}u_{e_{b,j}}
=
\Theta(u_{\max}).
\end{aligned}
\label{eq:tail_block_delta_scaling}
\end{equation}
Under the nonnegative block-correction condition above and for a fixed number of dependence blocks, all block corrections are nonnegative and of order at most \(u_{\max}\). Hence,
\begin{equation}
\Omega_j^{\mathrm{dep}}
=
\Theta(u_{\max}),
\qquad
\frac{
\Omega_j^{\mathrm{dep}}-\Omega_i^{\mathrm{dep}}
}{
P_1^{\mathrm{out}}
}
=
\Theta(1).
\label{eq:order_one_dependence_penalty}
\end{equation}
The normalized dependence contribution therefore remains of constant order and dominates the \(O(u_{\max})\) difference between the candidates' marginal outage probabilities. Lower-tail dependence increases the probability of simultaneous low-SNR events on the primary and backup paths. Because this penalty may persist at high average SNR, marginal-outage-based routing can overestimate the backup-path reliability gain.

Suppose further that both candidates are edge-disjoint from the primary path or, more generally, that
\begin{equation}
\frac{
\Omega_j^{\mathrm{sh}}-\Omega_i^{\mathrm{sh}}
}{
P_1^{\mathrm{out}}
}
=
o(1).
\label{eq:negligible_shared_difference}
\end{equation}
It then follows from \eqref{eq:general_score_difference} that
\begin{equation}
J_i-J_j
=
-\Theta(1)+o(1),
\label{eq:first_order_strict_separation}
\end{equation}
and therefore \(J_i<J_j\) for all sufficiently small \(u_{\max}\). Thus, backup-path selection may favor spatial, topological, or propagation diversity over marginal reliability alone, and a slightly less reliable candidate may be the better backup when it is less dependent on the primary path.
Finally, let
\begin{equation}
Q_k
\triangleq
\Pr\!\left(
\mathcal{S}_k^c\mid\mathcal{S}_1^c
\right)
=
J_k+O(A_k),
\qquad
A_k=O(u_{\max}).
\label{eq:exact_ranking_tail_agreement}
\end{equation}
For candidates \(i\) and \(j\),
\begin{equation}
\begin{aligned}
Q_i-Q_j
&=
J_i-J_j+O(A_i+A_j)
\\
&=
-\Theta(1)+o(1).
\end{aligned}
\end{equation}
Thus, \(Q_i<Q_j\) for all sufficiently small \(u_{\max}\). Accordingly, the exact conditional-outage metric and the
first-order score asymptotically rank \(\mathcal{P}_i\) ahead of \(\mathcal{P}_j\).

\section{Numerical Results}
\label{sec:simulation}

This section evaluates the proposed framework numerically and validates the analytical results using Monte Carlo simulations with \(N=10^6\) realizations. Rayleigh fading is assumed in the SNR-domain experiments, and pointwise \(95\%\) confidence intervals are omitted for visual clarity. Gaussian, Student-\(t\), Clayton, and Gumbel copulas are considered, with \(\nu=10\) for the Student-\(t\) copula. The copula parameters are determined using the parameter mappings based on Kendall's \(\tau\) in Table~\ref{tab:tau_relations}. When two paths share a physical link, the same link realization is used for both paths.

\begin{figure}[!t]
\centering
\includegraphics[width=0.93\columnwidth]{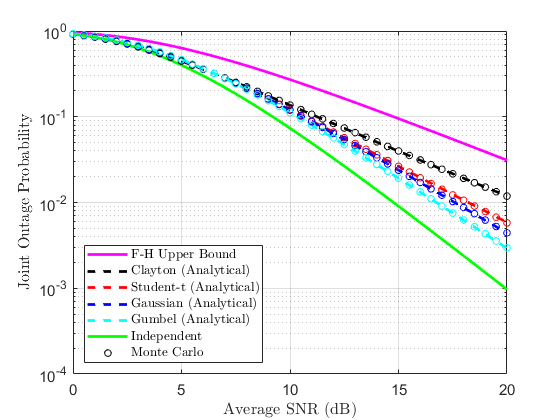}
\vspace{-2mm}
\caption{Joint outage probability for two dependent Rayleigh-fading links with \(\tau=0.25\) and \(\gamma^{\mathrm{th}}=5~\mathrm{dB}\).}
\label{fig:copula_outage}
\vspace{-3mm}
\end{figure}

Fig.~\ref{fig:copula_outage} considers two identically distributed Rayleigh-fading links forming one bivariate dependence block, with no shared links. The average SNR varies from \(0\) to \(20~\mathrm{dB}\), \(\gamma^{\mathrm{th}}=5~\mathrm{dB}\), and \(\tau=0.25\); the Student-\(t\) copula uses \(\nu=10\). The analytical and Monte Carlo results closely agree. At high SNR, the Clayton copula yields the largest outage probability because of its lower-tail dependence, followed by the Student-\(t\) copula. For the selected \(\tau\), SNR threshold, and average-SNR range, the Gaussian and Gumbel copulas produce smaller lower-tail joint outage probabilities.

\begin{figure}[t]
\centering
\includegraphics[width=0.93\columnwidth]{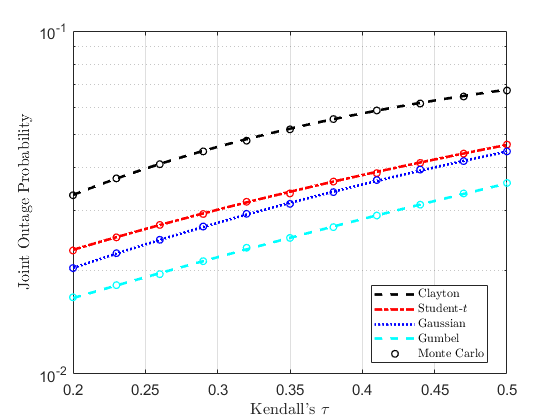}
\vspace{-2mm}
\caption{Joint outage probability versus Kendall's \(\tau\) for two Rayleigh-fading links with \(\bar{\gamma}=15~\mathrm{dB}\) and \(\gamma^{\mathrm{th}}=5~\mathrm{dB}\).}
\label{fig:rayleigh_tau}
\vspace{-3mm}
\end{figure}

Fig.~\ref{fig:rayleigh_tau} considers two identically distributed Rayleigh-fading links forming one bivariate dependence block, with no shared links. The average SNR is fixed at \(15~\mathrm{dB}\), \(\gamma^{\mathrm{th}}=5~\mathrm{dB}\), and Kendall's \(\tau\) varies from \(0.20\) to \(0.50\); the Student-\(t\) copula uses \(\nu=10\). The joint outage probability increases with dependence strength for all considered copulas, with the Clayton copula producing the largest outage probability at this operating point.

\begin{figure}[t]
\centering
\includegraphics[width=0.93\columnwidth]{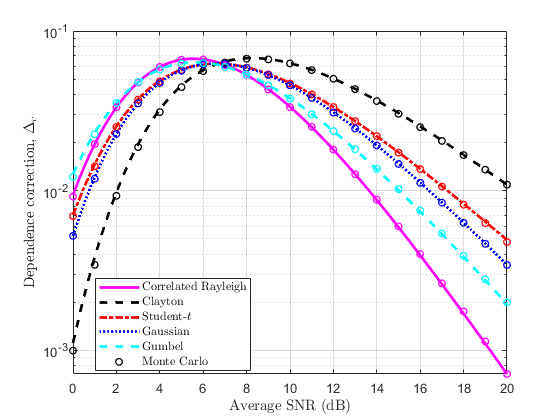}
\vspace{-2mm}
\caption{Dependence correction relative to the independent benchmark for models calibrated to the target Kendall's \(\tau=0.25\).}\label{fig:delta}
\vspace{-3mm}
\end{figure}

Fig.~\ref{fig:delta} considers two identically distributed Rayleigh-fading links forming one bivariate dependence block, with no shared links. The average SNR varies from \(0\) to \(20~\mathrm{dB}\), \(\gamma^{\mathrm{th}}=5~\mathrm{dB}\), \(\tau=0.25\), and the Student-\(t\) copula uses \(\nu=10\). The dependence correction is \(\Delta=C(u,u)-u^2\), where \(u\) is the marginal link-outage probability and \(C\) is the corresponding bivariate copula. The correction vanishes in both the low- and high-SNR regimes and reaches its maximum at an intermediate SNR. In the high-reliability regime, the Clayton copula yields the largest correction because of its positive lower-tail dependence.

For a fair comparison, the conventional correlated Rayleigh-fading benchmark is calibrated to the same Kendall's \(\tau\) as the copula models. Let \(X_i=|h_i|^2\), where \(h_1\) and \(h_2\) are jointly proper complex Gaussian channel coefficients with unit average powers. For each target \(\tau\), the power-correlation coefficient \(\rho_{\mathrm{power}}=\operatorname{corr}(X_1,X_2)=|\rho_h|^2\) is calibrated numerically by estimating Kendall's \(\tau\) from simulated power-gain samples and applying bisection. For Monte Carlo generation, \(\rho_h=\sqrt{\rho_{\mathrm{power}}}\) is taken to be real and nonnegative, with \(h_1=w_1\) and \(h_2=\rho_h h_1+\sqrt{1-\rho_{\mathrm{power}}}\,w_2\), where \(w_1,w_2\sim\mathcal{CN}(0,1)\) are independent.

\begin{figure}[t]
\centering
\includegraphics[width=0.93\columnwidth]{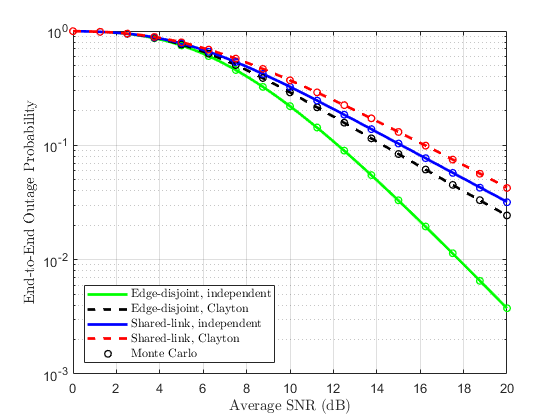}
\vspace{-2mm}
\caption{End-to-end outage for edge-disjoint and shared-link dual-path configurations with matched marginal link-outage probabilities.}
\label{fig:outage_copula_topology}
\vspace{-3mm}
\end{figure}

Fig.~\ref{fig:outage_copula_topology} compares two dual-path configurations, each containing two hops per path and having identical Rayleigh marginals. In the edge-disjoint case, \(E(\mathcal{P}_1)=\{e_1^{(1)},e_2^{(1)}\}\) and \(E(\mathcal{P}_2)=\{e_1^{(2)},e_2^{(2)}\}\), with \((e_1^{(1)},e_1^{(2)})\) and \((e_2^{(1)},e_2^{(2)})\) forming independent Clayton blocks with \(\tau=0.25\). In the shared-link case, \(E(\mathcal{P}_1)=\{e_1^{\mathrm{sh}},e_1^{(1)}\}\) and \(E(\mathcal{P}_2)=\{e_1^{\mathrm{sh}},e_1^{(2)}\}\). The shared edge is independent of both branch links; the branch links are either mutually independent or coupled through a Clayton copula with \(\tau=0.25\). All links use
\(\gamma^{\mathrm{th}}=5~\mathrm{dB}\), and their common average SNR varies from \(0\) to \(20~\mathrm{dB}\).

\begin{figure}[t]
\centering
\includegraphics[width=0.93\columnwidth]{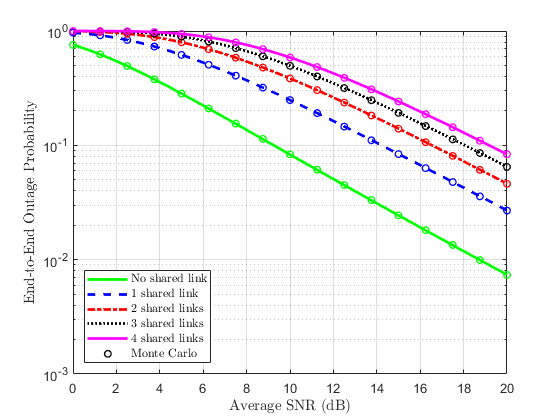}
\vspace{-2mm}
\caption{End-to-end outage versus the length of a common path segment, showing the combined effects of increasing path length and shared-link overlap.}
\label{fig:outage_clayton_shared_edges}
\vspace{-3mm}
\end{figure}

Fig.~\ref{fig:outage_clayton_shared_edges} considers \(E(\mathcal{P}_k)=\{e_1^{\mathrm{sh}},\ldots,e_r^{\mathrm{sh}}, e_1^{(k)}\}\), \(k\in\{1,2\}\), where \(r\in\{0,1,2,3,4\}\); the shared-edge sequence is empty for \(r=0\). Thus, each path has \(r+1\) hops. The shared edges are mutually independent and independent of the Clayton-dependent branch pair \((e_1^{(1)},e_1^{(2)})\), with \(\tau=0.25\). All links have identical Rayleigh marginals, use \(\gamma^{\mathrm{th}}=3~\mathrm{dB}\), and share an average SNR that varies from \(0\) to \(20~\mathrm{dB}\).

\begin{figure}[t]
\centering
\includegraphics[width=0.93\columnwidth]{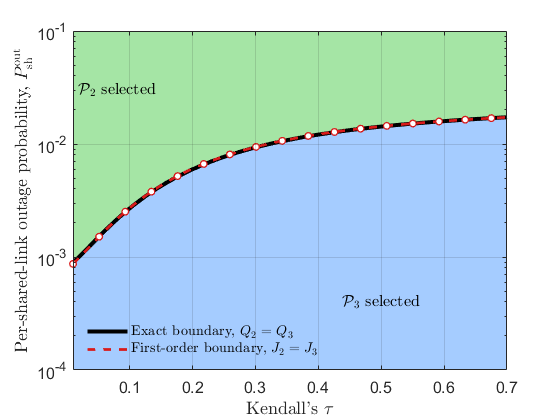}
\vspace{-2mm}
\caption{Backup-path selection regions versus Kendall's \(\tau\) and \(P_{\mathrm{sh}}^{\mathrm{out}}\). The solid and dashed curves denote the exact boundary \(Q_2=Q_3\) and the first-order boundary \(J_2=J_3\), respectively.}
\label{fig:backup_selection_region}
\vspace{-3mm}
\end{figure}

Fig.~\ref{fig:backup_selection_region} compares two candidate backup paths under a common block-factorized model. Let
\(E(\mathcal{P}_1)=\{e_1,\ldots,e_6\}\),
\(E(\mathcal{P}_2)=\{e_1,e_7,e_8,e_9\}\), and
\(E(\mathcal{P}_3)=\{e_1,e_2,e_{10}\}\). Hence,
\(E_{12}^{\mathrm{sh}}=\{e_1\}\),
\(E_{13}^{\mathrm{sh}}=\{e_1,e_2\}\),
\(\mathcal{D}_{12}=\{(e_3,e_7),(e_4,e_8),(e_5,e_9)\}\), and
\(\mathcal{D}_{13}=\{(e_6,e_{10})\}\).
All dependence blocks, shared links, and remaining links are mutually independent, and the dependence within each block is modeled by a Clayton copula with the same Kendall's \(\tau\). Each non-shared link has outage probability \(u=0.01\), and each shared link has outage probability
\(p=P_{\mathrm{sh}}^{\mathrm{out}}\). We vary
\(\tau\in[0.01,0.70]\) and \(p\in[10^{-4},10^{-1}]\). Hence,
\begin{equation}
\begin{aligned}
P_1^{\mathrm{out}}&=1-(1-p)^2(1-u)^4,\\
P_2^{\mathrm{out}}&=1-(1-p)(1-u)^3,\\
P_3^{\mathrm{out}}&=1-(1-p)^2(1-u).
\end{aligned}
\end{equation}
For each dependence block,
\begin{equation}
\Delta(\tau)
=
\left(2u^{-\theta(\tau)}-1\right)^{-1/\theta(\tau)}-u^2,
\qquad
\theta(\tau)=\frac{2\tau}{1-\tau}.
\end{equation}
The first-order score difference is
\begin{equation}
J_2-J_3
=
P_2^{\mathrm{out}}-P_3^{\mathrm{out}}
+
\frac{2\Delta(\tau)-p}{P_1^{\mathrm{out}}}.
\end{equation}
The first-order and exact boundaries are determined by \(J_2=J_3\) and \(Q_2=Q_3\), respectively. The two selection rules identify the same backup path for \(99.91\%\) of the points on the \(250\times250\) evaluation grid.

\begin{figure}[t]
\centering
\includegraphics[width=0.93\columnwidth]{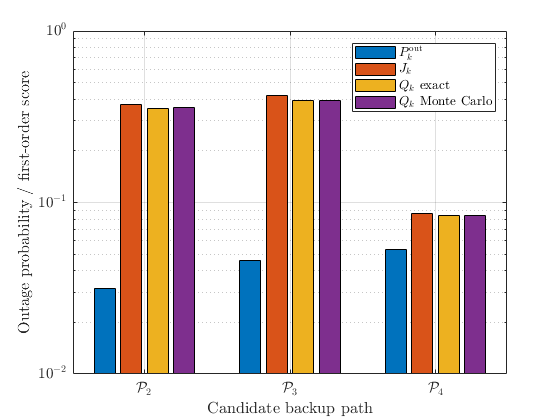}
\vspace{-2mm}
\caption{Marginal outage probability, first-order score, exact conditional outage probability, and Monte Carlo estimate for the candidate backup paths.}
\label{fig:backup_comparison}
\vspace{-3mm}
\end{figure}

Fig.~\ref{fig:backup_comparison} compares the candidate backup paths for the topology in Fig.~\ref{fig:network}, where the four-hop path \(\mathcal{P}_1\), whose edge set is \(E(\mathcal{P}_1)=\{e_1,e_2,e_3,e_4\}\), is designated as the primary path. The candidate paths, dependence blocks, shared links, and pairwise Kendall's \(\tau\) values are listed in Table~\ref{tab:backup_repro}. All dependence blocks use Clayton copulas. Distinct dependence blocks and all other links are mutually independent, while the same realization of the shared link \(e_4\) is used for \(\mathcal{P}_1\) and \(\mathcal{P}_3\).

Selection based only on marginal outage favors \(\mathcal{P}_2\), but the strong dependence between \(\mathcal{P}_2\) and the primary path substantially increases the conditional outage of \(\mathcal{P}_2\). Candidate \(\mathcal{P}_3\) is additionally penalized by the shared link \(e_4\). Although \(\mathcal{P}_4\) has a higher marginal outage probability than \(\mathcal{P}_2\), its weaker dependence and absence of shared links yield the lowest conditional outage probability. Thus, the first-order selection rule, the exact conditional-outage rule, and the Monte Carlo simulation all identify $\mathcal{P}_4$ as the preferred backup path.

\begin{figure}[t]
\centering
\includegraphics[width=0.93\columnwidth]{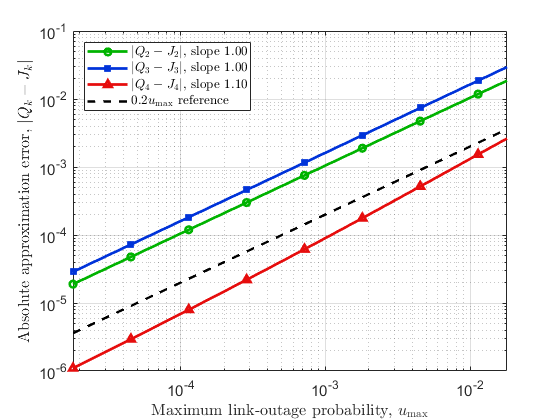}
\vspace{-2mm}
\caption{Absolute approximation error
\(\lvert Q_k-J_k\rvert\) versus \(u_{\max}\) on logarithmic axes. The dashed line has unit slope and is included as a reference for first-order scaling.}
\label{fig:first_order_error}
\vspace{-3mm}
\end{figure}

The link-outage probabilities are specified directly as
\begin{equation}
\mathbf{u}
=
10^{-3}
[15,\,10,\,12,\,18,\,10,\,12,\,10,\,9,\,11,\,16,\,15,\,14,\,9],
\end{equation}
which gives \(P_1^{\mathrm{out}}=0.053894\). The Monte Carlo simulation uses \(N=10^6\) realizations with a random seed of \(10\). Because Figs.~\ref{fig:backup_selection_region},~\ref{fig:backup_comparison}, and~\ref{fig:first_order_error} are parameterized directly by link-outage probabilities, average SNRs and SNR thresholds are not used in these experiments.

\begin{table}[t]
\centering
\caption{Parameters and results for
Fig.~\ref{fig:backup_comparison}.}
\label{tab:backup_repro}
\resizebox{\columnwidth}{!}{%
\begin{tabular}{lccc}
\hline
 & \(\mathcal{P}_2\) & \(\mathcal{P}_3\) & \(\mathcal{P}_4\) \\
\hline
Links
& \(e_5,e_6,e_7\)
& \(e_8,e_9,e_{13},e_4\)
& \(e_8,e_{10},e_{11},e_{12}\) \\

Number of hops
& \(3\) & \(4\) & \(4\) \\

Shared links
& None & \(e_4\) & None \\

Dependent pairs
& \((e_1,e_5),(e_4,e_7)\)
& \((e_1,e_8),(e_3,e_{13})\)
& \((e_1,e_8),(e_4,e_{12})\) \\

Pairwise \(\tau\)
& \((0.60,0.60)\)
& \((0.10,0.10)\)
& \((0.10,0.05)\) \\

\(P_k^{\mathrm{out}}\)
& \(0.0317\)
& \(0.0462\)
& \(0.0529\) \\

\(\Omega_k^{\mathrm{sh}}\)
& \(0\)
& \(0.0180\)
& \(0\) \\

\(\Omega_k^{\mathrm{dep}}\)
& \(0.01833\)
& \(0.00217\)
& \(0.00179\) \\

\(J_k\)
& \(0.3717\)
& \(0.4204\)
& \(0.0862\) \\

\(Q_k\)
& \(0.3532\)
& \(0.3909\)
& \(0.0835\) \\
\hline
\end{tabular}}
\vspace{-2mm}
\end{table}

Finally, Fig.~\ref{fig:first_order_error} evaluates the accuracy of the first-order score using the topology, baseline link-outage vector, Clayton dependence blocks, and pairwise Kendall's $\tau$ values shown in Fig.~\ref{fig:backup_comparison} and listed in Table~\ref{tab:backup_repro}. We scale all link-outage probabilities as
\(u_e(\varepsilon)=\varepsilon u_e^{(0)}\), with
\(\varepsilon\in[10^{-3},1]\), while keeping the topology and copula parameters fixed. Hence,
\(u_{\max}=\varepsilon\max_e u_e^{(0)}\), which tends to zero as \(\varepsilon\to0\). For each candidate, the linear model
\begin{equation}
\log_{10}\lvert Q_k-J_k\rvert
=
a_k\log_{10}u_{\max}+b_k
\end{equation}
is fitted over \(u_{\max}\leq2\times10^{-3}\). The fitted slopes are approximately \(1.00\), \(1.00\), and \(1.10\)
for \(\mathcal{P}_2\), \(\mathcal{P}_3\), and \(\mathcal{P}_4\),
respectively. These numerical results illustrate the first-order error
scaling established in Proposition~\ref{prop:first_order_conditional_score}.

\section{Conclusion}
\label{sec:conclusion}

This paper developed a copula-based framework for analyzing dependent outages in primary--backup multi-hop wireless networks. Under the adopted pairwise block-factorized model, exact joint and conditional outage expressions were derived for edge-disjoint and overlapping paths. The framework explicitly distinguishes between statistical dependence among distinct cross-path links and common failures caused by physically shared links.

In the high-reliability regime, a first-order backup-path score was obtained by separating the effects of marginal outage, shared links, and node-induced dependence. Sufficient conditions for agreement between the approximate and exact rankings were also established. The resulting score provides an interpretable criterion for dependence-aware backup-path selection.

The analysis and numerical examples show that tail dependence and path overlap can reduce redundancy gains and reverse rankings based only on marginal path outage. In particular, copulas with the same Kendall's \(\tau\) may produce markedly different conditional-outage behavior. 

\appendices

\section{Proof of Proposition~\ref{prop:exact_dual_path_outage}}
\label{app:exact_dual_path_outage}
For \(j\in\{1,k\}\), let
\[
A_j=\{\gamma_{a_j}<\gamma_{a_j}^{\mathrm{th}}\},
\qquad
D_j=\{\gamma_{b_j}<\gamma_{b_j}^{\mathrm{th}}\}
\]
denote the outage events of the first and last links of
\(\mathcal{P}_j\), respectively, and define the intermediate-link
outage event as
\[
I_j
=
\bigcup_{i=1}^{L_j-2}
\left\{
\gamma_{e_i^{(j)}}
<
\gamma_{e_i^{(j)}}^{\mathrm{th}}
\right\}.
\]
For \(L_j=2\), the intermediate segment is empty and
\(\Pr(I_j^c)=1\); otherwise, independence of the intermediate
links gives
\[
\Pr(I_j^c)
=
\prod_{i=1}^{L_j-2}
\left(1-u_{e_i^{(j)}}\right)
=
\beta_j.
\]

The path-success event is
\(\mathcal{S}_j=A_j^c\cap I_j^c\cap D_j^c\). Hence, by
inclusion--exclusion,
\begin{equation}
\begin{aligned}
P_{1k}^{\mathrm{out}}
&=
\Pr\!\left(
\mathcal{S}_1^c\cap\mathcal{S}_k^c
\right)
\\
&=
1-\Pr(\mathcal{S}_1)-\Pr(\mathcal{S}_k)
+\Pr(\mathcal{S}_1\cap\mathcal{S}_k).
\end{aligned}
\label{eq:proof_pairwise_inclusion_exclusion}
\end{equation}

From the definition of \(\Delta_s\),
\[
\Pr(A_1\cap A_k)
=
C_{E_s^{\mathrm{out}}}(u_{a_1},u_{a_k})
=
u_{a_1}u_{a_k}+\Delta_s.
\]
Therefore,
\[
\Pr(A_1^c\cap A_k^c)
=
(1-u_{a_1})(1-u_{a_k})+\Delta_s.
\]
Similarly,
\[
\Pr(D_1^c\cap D_k^c)
=
(1-u_{b_1})(1-u_{b_k})+\Delta_d.
\]

Under the adopted dependence model, the source-side, destination-side, and intermediate-link groups are mutually independent. Hence,
\[
\Pr(\mathcal{S}_j)
=
(1-u_{a_j})\beta_j(1-u_{b_j}),
\qquad j\in\{1,k\},
\]
and
\begin{equation}
\begin{aligned}
\Pr(\mathcal{S}_1\cap\mathcal{S}_k)
&=
\beta_1\beta_k
\left[
(1-u_{a_1})(1-u_{a_k})+\Delta_s
\right]
\\
&\quad\times
\left[
(1-u_{b_1})(1-u_{b_k})+\Delta_d
\right].
\end{aligned}
\label{eq:proof_joint_path_success}
\end{equation}

For the path-independent benchmark, the marginal path-success probabilities remain unchanged, while the endpoint dependence corrections are removed. Therefore,
\begin{equation}
\begin{aligned}
P_{1k,\mathrm{pind}}^{\mathrm{out}}
&=
P_1^{\mathrm{out}}P_k^{\mathrm{out}}
\\
&=
1-\Pr(\mathcal{S}_1)-\Pr(\mathcal{S}_k)
\\
&\quad+
\beta_1\beta_k
(1-u_{a_1})(1-u_{a_k})
(1-u_{b_1})(1-u_{b_k}).
\end{aligned}
\label{eq:proof_pairwise_independent_benchmark}
\end{equation}

Subtracting
\eqref{eq:proof_pairwise_independent_benchmark} from
\eqref{eq:proof_pairwise_inclusion_exclusion} and expanding
\eqref{eq:proof_joint_path_success} gives
\begin{equation}
\begin{aligned}
P_{1k}^{\mathrm{out}}
-
P_{1k,\mathrm{pind}}^{\mathrm{out}}
&=
\beta_1\beta_k
\Big[
\Delta_s(1-u_{b_1})(1-u_{b_k})
\\
&\quad+
\Delta_d(1-u_{a_1})(1-u_{a_k})
+
\Delta_s\Delta_d
\Big].
\end{aligned}
\label{eq:exact_dual_path_outage_correction}
\end{equation}
This proves \eqref{eq:exact_dual_path_outage_decomposition}.

\section{Proof of Corollary~\ref{cor:high_reliability_dual_path}}

Because the path lengths are finite and \(u_e\leq u_{\max}\), we have
\(\beta_1\beta_k=1+O(u_{\max})\). Moreover, both
\((1-u_{b_1})(1-u_{b_k})\) and
\((1-u_{a_1})(1-u_{a_k})\) are \(1+O(u_{\max})\).
Substituting these relations into
\eqref{eq:exact_dual_path_outage_decomposition} gives
\begin{equation}
\begin{aligned}
P_{1k}^{\mathrm{out}}
&=
P_{1k,\mathrm{pind}}^{\mathrm{out}}
+
\Delta_s+\Delta_d
\\
&\quad+
O\!\left(
u_{\max}(\Delta_s+\Delta_d)
+
\Delta_s\Delta_d
\right).
\end{aligned}
\end{equation}

Under positive quadrant dependence,
\(\Delta_s\geq0\) and \(\Delta_d\geq0\).
Moreover, the Fr\'echet--Hoeffding upper bound gives
\(0\leq\Delta_s\leq u_{\max}\) and
\(0\leq\Delta_d\leq u_{\max}\). Hence,
\[
\Delta_s\Delta_d
\leq
u_{\max}\min(\Delta_s,\Delta_d)
\leq
u_{\max}(\Delta_s+\Delta_d),
\]
and consequently
\begin{equation}
P_{1k}^{\mathrm{out}}
=
P_{1k,\mathrm{pind}}^{\mathrm{out}}
+
\Delta_s+\Delta_d
+
O\!\left(
u_{\max}(\Delta_s+\Delta_d)
\right).
\end{equation}
Thus, to leading order,
\[
P_{1k}^{\mathrm{out}}
\approx
P_{1k,\mathrm{pind}}^{\mathrm{out}}
+
\Delta_s+\Delta_d.
\]

Using
\[
\Delta_{k\mid1}
=
\frac{
P_{1k}^{\mathrm{out}}
-
P_1^{\mathrm{out}}P_k^{\mathrm{out}}
}{
P_1^{\mathrm{out}}
},
\]
together with
\(P_{1k,\mathrm{pind}}^{\mathrm{out}}
=P_1^{\mathrm{out}}P_k^{\mathrm{out}}\), we obtain, for
\(\Delta_s+\Delta_d>0\),
\[
\begin{aligned}
\Delta_{k\mid1}
&=
\frac{
\Delta_s+\Delta_d
+
O\!\left(
u_{\max}(\Delta_s+\Delta_d)
\right)
}{
P_1^{\mathrm{out}}
}
\\
&=
\frac{\Delta_s+\Delta_d}{P_1^{\mathrm{out}}}
\left[1+O(u_{\max})\right].
\end{aligned}
\]
Hence,
\[
\Delta_{k\mid1}
\approx
\frac{\Delta_s+\Delta_d}{P_1^{\mathrm{out}}}.
\]
If \(\Delta_s=\Delta_d=0\), then
\eqref{eq:conditional_outage_gain_exact} directly gives
\(\Delta_{k\mid1}=0\).

\section{Proof of Proposition~\ref{prop:first_order_conditional_score}}
\label{app:first_order_conditional_score}

From \eqref{eq:exact_conditional_outage_blocks},
\begin{equation}
\Pr\!\left(
\mathcal{S}_k^c\mid\mathcal{S}_1^c
\right)
=
P_k^{\mathrm{out}}
+
\frac{\zeta_1\zeta_k}{P_1^{\mathrm{out}}}
\left(
\Lambda_{1k}^{\mathrm{sh}}
\Lambda_{1k}^{\mathrm{dep}}
-1
\right).
\label{eq:proof_exact_conditional}
\end{equation}

For every shared link \(e\in E_{1k}^{\mathrm{sh}}\),
\((1-u_e)^{-1}=1+u_e+O(u_e^2)\). Since the number of shared links remains fixed and \(u_e\leq u_{\max}\), the expansion
of the finite product in \eqref{eq:exact_shared_factor} gives
\begin{equation}
\Lambda_{1k}^{\mathrm{sh}}
=
1+\Omega_k^{\mathrm{sh}}
+
O\!\left(
u_{\max}\Omega_k^{\mathrm{sh}}
\right).
\label{eq:proof_shared_factor_expansion}
\end{equation}

For each dependence block \(b\in\mathcal{D}_{1k}\), the nonnegative block-correction condition above and the Fr\'echet--Hoeffding upper bound imply
\begin{equation}
0
\leq
\Delta_{b,k}
\leq
\min\!\left(
u_{e_{b,1}},u_{e_{b,k}}
\right)
-
u_{e_{b,1}}u_{e_{b,k}}
\leq
u_{\max}.
\label{eq:proof_delta_bound}
\end{equation}
Moreover,
\begin{equation}
\begin{aligned}
\frac{\Delta_{b,k}}
{(1-u_{e_{b,1}})(1-u_{e_{b,k}})}
&=
\Delta_{b,k}\left[1+O(u_{\max})\right]
\\
&=
\Delta_{b,k}
+
O\!\left(u_{\max}\Delta_{b,k}\right).
\end{aligned}
\label{eq:proof_normalized_delta_expansion}
\end{equation}
Because the number of dependence blocks remains fixed, expanding
the finite product in \eqref{eq:exact_dependence_factor} gives
\begin{equation}
\Lambda_{1k}^{\mathrm{dep}}
=
1+\Omega_k^{\mathrm{dep}}
+
O\!\left(
u_{\max}\Omega_k^{\mathrm{dep}}
\right).
\label{eq:proof_dependence_factor_expansion}
\end{equation}

Under the nonnegative block-correction condition,
\(\Omega_k^{\mathrm{sh}},\Omega_k^{\mathrm{dep}}\geq0\).
Combining \eqref{eq:proof_shared_factor_expansion} and
\eqref{eq:proof_dependence_factor_expansion}, and using
\(A_k=\Omega_k^{\mathrm{sh}}+\Omega_k^{\mathrm{dep}}\), yields
\begin{equation}
\Lambda_{1k}^{\mathrm{sh}}
\Lambda_{1k}^{\mathrm{dep}}
-1
=
A_k
+
O\!\left(
u_{\max}A_k
\right).
\label{eq:proof_combined_factor_expansion}
\end{equation}
Indeed,
\(\Omega_k^{\mathrm{sh}}\Omega_k^{\mathrm{dep}}
=O(u_{\max}A_k)\), because both aggregate contributions are
\(O(u_{\max})\).

Furthermore, finite path lengths and \(u_e\leq u_{\max}\) imply
\(P_j^{\mathrm{out}}=O(u_{\max})\) for \(j\in\{1,k\}\).
Hence,
\(\zeta_j=1-P_j^{\mathrm{out}}=1+O(u_{\max})\), and consequently
\begin{equation}
\zeta_1\zeta_k
=
1+O(u_{\max}).
\label{eq:proof_marginal_success_expansion}
\end{equation}

Substituting \eqref{eq:proof_combined_factor_expansion} and
\eqref{eq:proof_marginal_success_expansion} into
\eqref{eq:proof_exact_conditional} yields
\begin{equation}
\Pr\!\left(
\mathcal{S}_k^c\mid\mathcal{S}_1^c
\right)
=
P_k^{\mathrm{out}}
+
\frac{A_k}{P_1^{\mathrm{out}}}
+
O\!\left(
\frac{u_{\max}A_k}{P_1^{\mathrm{out}}}
\right).
\label{eq:proof_conditional_first_order}
\end{equation}

Finally, since \(P_1^{\mathrm{out}}=\Theta(u_{\max})\),
\[
O\!\left(
\frac{u_{\max}A_k}{P_1^{\mathrm{out}}}
\right)
=
O(A_k).
\]
Therefore,
\[
\Pr\!\left(
\mathcal{S}_k^c\mid\mathcal{S}_1^c
\right)
=
P_k^{\mathrm{out}}
+
\frac{
\Omega_k^{\mathrm{sh}}
+
\Omega_k^{\mathrm{dep}}
}{
P_1^{\mathrm{out}}
}
+
O(A_k),
\]
which proves \eqref{eq:conditional_outage_score_expansion}.

\section*{Acknowledgment}
This work was partially supported by Research Council of Finland funded
projects ReWin-6G (Grant Number 357120), 6G-Concorse (Grant Number 359850)
and 6G Flagship (Grant Number 369116).
\bibliographystyle{IEEEtran}
\bibliography{paper}

\end{document}